\documentclass[aps,prb,twocolumn,showpacs]{revtex4-1}
\usepackage{graphicx}% Include figure files
\usepackage{natbib}
\usepackage[utf8]{inputenc}
\usepackage{epstopdf}
\setcitestyle{super}
\usepackage{color}
\usepackage[normalem]{ulem}
\usepackage{pdfpages}
\makeatletter
\AtBeginDocument{\let\LS@rot\@undefined}
\makeatother

\newcommand*{\citen}[1]{%
  \begingroup
    \romannumeral-`\x % remove space at the beginning of \setcitestyle
    \setcitestyle{numbers}%
    \cite{#1}%
  \endgroup   
}

\usepackage{hyphenat}
\begin{document}

\preprint{APS/123-QED}
\title{Role of Fermi surface for the pressure-tuned nematic transition in the BaFe$_2$As$_2$ family}
\author{Elena Gati$^{1,2}$}
\author{Li Xiang$^{1,2}$}
\author{Sergey L. Bud'ko$^{1,2}$}
\author{Paul C. Canfield$^{1,2}$}

\address{$^{1}$ Ames Laboratory, US Department of Energy, Iowa State University, Ames,
Iowa 50011, USA}
\address{$^{2}$ Department of Physics and Astronomy, Iowa State University, Ames, Iowa 50011,
USA}

\date{\today}

\begin{abstract}
The tetragonal-to-orthorhombic phase transition at $T_s$, which precedes the antiferromagnetic phase transition at $T_N$ in many iron-based superconductors, is considered as one of the manifestations of electronic \textit{nematic} order. By constructing temperature-pressure phase diagrams of pure and Co-doped BaFe$_2$As$_2$, we study the relation of $T_s$ and $T_N$ under pressure $p$. Our data disclose two qualitatively different regimes in which $\Delta T\,=\,T_s-T_N$ either increases or decreases with $p$. We provide experimental evidence that the transition between the two regimes may be associated with sudden changes of the Fermi surface topology. Therefore, our results not only support the electronic origin of the structural order, but also emphasize the importance of details of the Fermi surface for the evolution of nematic order under pressure.
\end{abstract}

\pacs{xxx}

\maketitle

\textit{Introduction-} The family of iron-based superconductors manifests rich phase diagrams, which reflect a complex interplay of structure, magnetism and superconductivity \cite{Paglione10,Stewart11}. Most of the parent compounds in this family undergo a magnetic transition to a stripe-type antiferromagnetic (afm) phase \cite{Rotter08,Johnston10,Dai15} at $T_{N}$, which is either accompanied or preceded by a structural transition from a high-temperature tetragonal to a low-temperature orthorhombic structure at $T_{s}\,\geq\,T_{N}$. Upon substitution \cite{Ni08,Canfield09,Avci12,Luetkens09,Rotundu09,Canfield10} or pressurization \cite{Colombier09,Kim11b,Wu13,Kimber09}, both transitions are typically suppressed in temperature, while still closely following each other and superconductivity emerges \cite{Sefat08,Takahashi08,Rotter08,Alireza09} above a certain material-specific substitution level or pressure. The close relation of superconductivity with magnetic and structural degrees of freedom \cite{Hirschfeld11,Chubukov12,Si16} has raised significant interest in understanding the origin and interplay of the magnetic and structural transitions in these compounds. Nowadays, in particular in cases of $T_s\,>\,T_N$, the structural transition is considered as one of the manifestations of a nematic phase \cite{Fernandes14,Fernandes10,Fradkin10}, i.e., a phase with broken $C_4$ tetragonal symmetry, but preserved $O(3)$ spin-rotational symmetry, that precedes magnetic order. In fact, nematic order is suggested to be a prime example of a more general concept of vestigial order, which is induced by fluctuations of a multi-component primary order \cite{Fernandes19} and potentially relevant in various classes of superconductors \cite{Fradkin15,Hecker18}. 

 Even though there is consensus that nematicity in the iron-based superconductors is driven by electronic degrees of freedom \cite{Chu10,Tanatar10,Nandi10,Fernandes10,Chu12,Boehmer14}, the microscopic origin of the electronic nematic phase, however, is still under debate: stripe-type afm fluctuations \cite{Fang08,Xu08,Fernandes10,Si08,Fernandes12,Capati11,Wysocki11,Hu12} and orbital fluctuations between Fe 3d orbitals \cite{Krueger09,Lee09,Yin10,Lv10,Chen10,Applegate12} have been suggested as possible driving forces of the nematic phase. As both of these types of orders break the same symmetry, a distinction between the two scenarios can likely only be achieved by a comparison of experimental results and microscopic modeling. As a consequence, a thorough explanation of what controls the extent in temperature of the purely nematic phase is lacking. In this context, a very peculiar example is FeSe \cite{Boehmer17} in which nematic order is observed without any indications for long-range magnetic order.

The missing link for the understanding and controlling nematic order might be unraveled by extensive experimental studies of the phase diagrams, which depict the evolution of long-range nematic and magnetic order, as a function of various external control parameters. So far, the following archetypical iron-pnictide phase diagram was shaped by investigations on the 122 family, in particular BaFe$_2$As$_2$, mostly using chemical substitution as a tuning parameter. The parent compound BaFe$_2$As$_2$ undergoes a second-order structural transition at $T_s\,\approx\,135$\,K, closely followed by a weak first-order magnetic phase transition at $T_N$ upon cooling \cite{Rotter08,Kim11,Rotundu11} ($\Delta T\,=\,T_s-T_N\,\approx\,$0.5\,K to 1\,K). For electron doping, the afm and structural transition temperatures rapidly split further upon increasing doping level with $T_s\,>\,T_N$ \cite{Canfield09,Canfield10,Kim11,Rotundu11}. Thus, the phase diagram upon electron doping reveals a wide region of purely nematic order. In contrast, no splitting of $T_s$ and $T_N$ was observed upon hole doping \cite{Avci12}.

		\begin{figure*}
		\begin{center}
		\includegraphics[width=\textwidth]{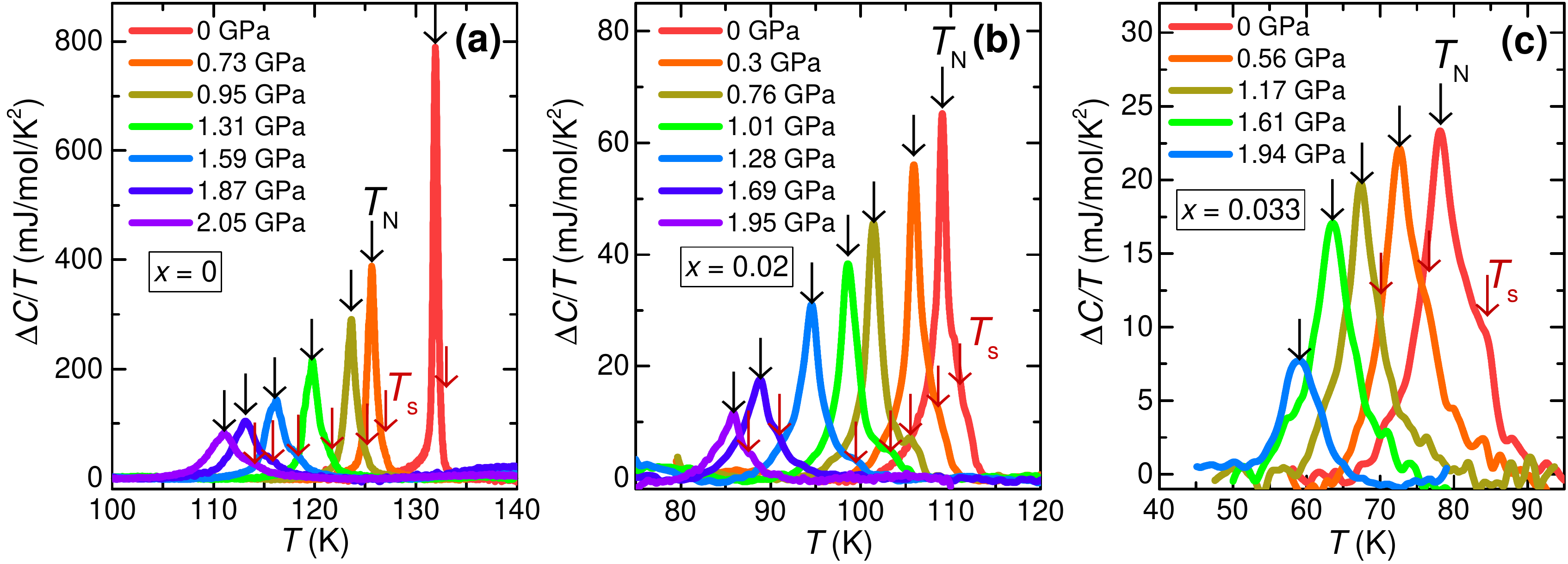} 
		\caption{Specific heat anomaly, $\Delta C/T$, vs. temperature $T$ of Ba(Fe$_{1-x}$Co$_x$)$_2$As$_2$ ($x\,=\,0$ (a), $x\,=\,0.02$ (b), $x\,=\,0.33$ (c)) at different pressures up to 2.05\,GPa. Black (brown) arrows mark the position of the inferred antiferromagnetic (structural) transition temperature at $T_N$ ($T_s$).}
		\label{fig:specanomaly}
		\end{center}
		\end{figure*}

Several pressure studies revealed a suppression of both $T_s$ and $T_N$\cite{Colombier09,Fukazawa08,Yamazaki10,Matsubayashi09,Ishikawa09}, which led to the common belief that pressure and electron doping can be considered as similar tuning parameters. However, no systematic experimental study has been conducted yet\cite{comment1, Wu13, Ikeda18} as to how $T_s$ is related to $T_N$ in the BaFe$_2$As$_2$ family under pressure, $p$. The tuning by pressure here is not only complementary to substitution studies, but also lacks any complications related to changes in substitution-induced disorder. Pressure-dependent studies of the phase diagram might therefore be considered as an even more clear example case for comparison with microcoscopic theories. By measuring the specific heat of the Ba(Fe$_{1-x}$Co$_x$)$_2$As$_2$ family under pressure, we are able to trace the magnetic and structural transition temperatures as a function of pressure by performing one single experiment (see Appendix Figs.\,\ref{fig:RTP}-\ref{fig:RTP-PD} for supporting transport data) in an unambiguous manner (without the need of invoking a timescale of our measurement technique). Our results clearly reveal the existence of two qualitatively different regimes in which $\Delta T\,=T_s-T_N$ either increases or decreases with $p$. By combining this data with Hall measurements under pressure, we assign the different response of the structural (= \textit{nematic}) transition to pressure to the system undergoing a sudden change of the Fermi surface topology. Thus, our results support the electronic origin of nematicity in the BaFe$_2$As$_2$ family and, even further, provide evidence that details of the Fermi surface play a key role for the pressure-tuned nematic transition. In turn, this allows us to discuss implications on electronic parameters governing the phase diagram of the 122 pnictides.

\textit{Methods-} As-grown single crystals of Ba(Fe$_{1-x}$Co$_x$)$_2$As$_2$ ($x\,=\,0,\, 0.02,\,0.033$) used in this study were grown out of self-flux, as described elsewhere\cite{Ni08}. Given $x$ values correspond to the measured rather than the nominal values \cite{Ni08}. Specific heat under pressure measurements were performed in a piston-cylinder pressure cell with maximum pressure of $\approx\,2\,$GPa using the AC calorimetry technique, as described in detail in Ref.\,\citen{Gati18}. Hall effect measurements under pressure were performed using the same procedure, as described in Ref.\,\citen{Mun09} for ambient $p$ studies (see SI). In both cases, a 4:6 mixture of light mineral oil and n-pentane was used as a pressure-transmitting medium. This pressure medium solidifies between 3 and 4\,GPa at room temperature \cite{Kim11b,Torikachvili15}, which ensures good hydrostaticity of the pressure environment in our experiment. Pressure values given in the entire manuscript correspond to the ones determined at low temperatures via the shift of the critical temperature of elemental lead \cite{Bireckoven88}.

\textit{Experimental Results-}	Figure \ref{fig:specanomaly} summarizes results of our pressure($p$)-dependent specific heat ($C$) study on in total three members of the Ba(Fe$_{1-x}$Co$_x$)$_2$As$_2$ family ($x\,=\, 0$ (a), 0.02 (b) and 0.033 (c)). The anomalous contributions to the specific heat, $\Delta C(T)/T$, which were obtained after subtraction of a background contribution (see SI Figs.\,\ref{fig:Ba122}-\ref{fig:BaCo122-2}), all reveal very similar features. We find a sharp peak in $\Delta C/T$ at all pressures, which becomes strongly reduced in size and shifts to lower temperature upon increasing $p$. As known from detailed ambient-$p$ thermodynamic and scattering studies \cite{Ni08,Chu09,Kim11}, this sharp peak in $\Delta C/T$ (indicated by the black arrows in Fig.\,\ref{fig:specanomaly}) corresponds to the signature of the afm phase transition at $T_N$. At the same time, depending on the separation of the structural and afm transition, most of the data sets reveal either a shoulder or an additional peak on the high-$T$ side of the afm peak (visualized by the brown arrows). This feature is known to be associated with the structural phase transition at $T_s$, again from ambient-$p$ thermodynamic and scattering studies \cite{Ni08,Chu09,Kim11}.
	
\begin{figure}
		\includegraphics[width=\columnwidth]{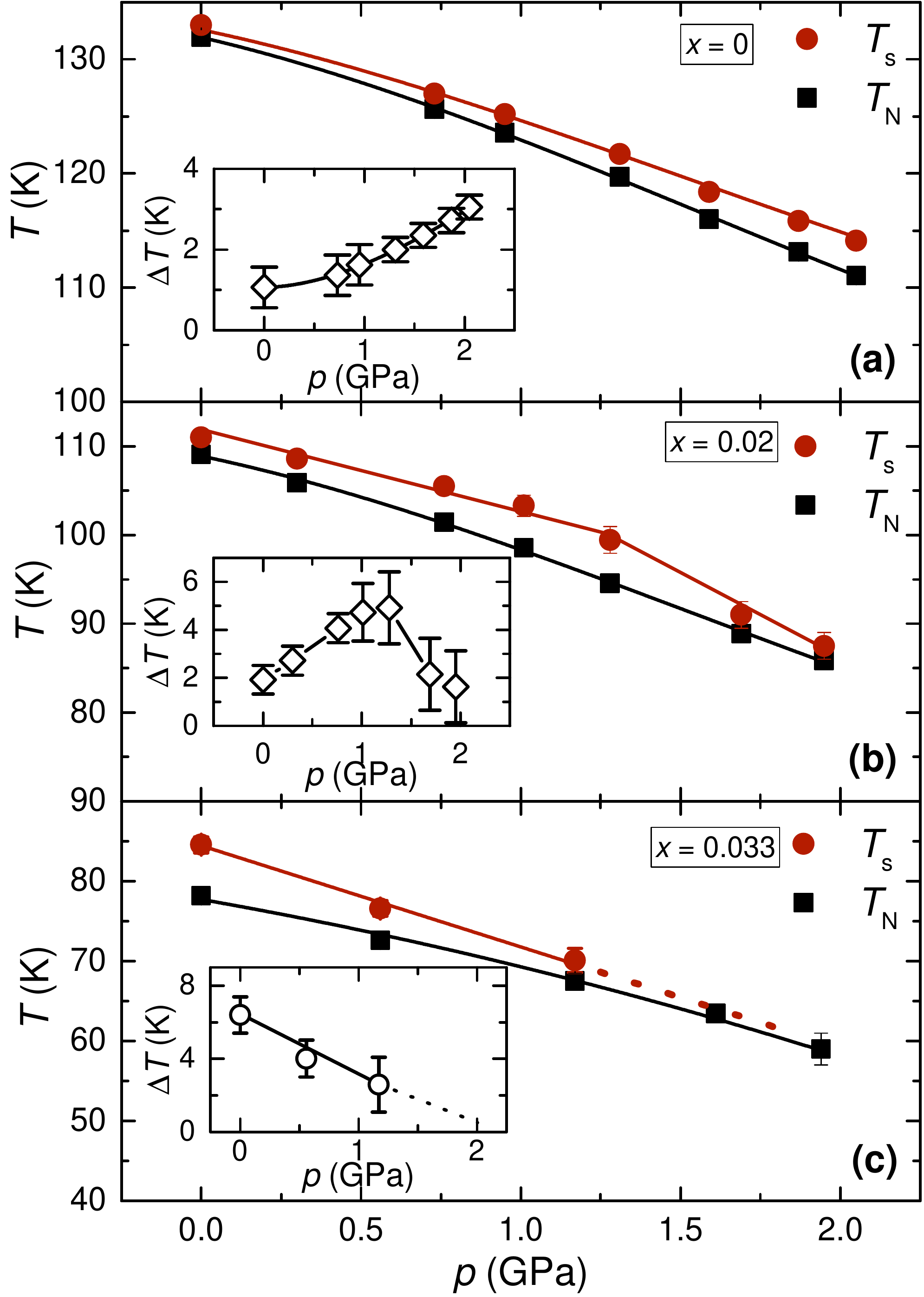} 
		\caption{Temperature ($T$)-pressure ($p$) phase diagram for Ba(Fe$_{1-x}$Co$_x$)$_2$As$_2$ (($x\,=\,0$ (a), $x\,=\,0.02$ (b), $x\,=\,0.33$ (c)). Black squares (brown circles) correspond to antiferromagnetic (structural) transition temperature at $T_N$ ($T_s$). Insets: Pressure dependence of $\Delta T\,=\,T_s-T_N$. Error bars in the main panel are smaller than the symbol size due to the large temperature scale.}
		\label{fig:phasediagram}
		\end{figure}

	By defining criteria based on the $C/T$ data sets, as well their $T$-derivatives (see SI Figs.\,\ref{fig:Ba122}-\ref{fig:criteriondiscussion3}), we construct $T$-$p$ phase diagrams, shown in Fig.\,\ref{fig:phasediagram}. These phase diagrams contain the main findings on the $p$ evolution of $T_N$ and $T_s$ of the present work. For all three studied compounds, we find a suppression of $T_s$ and $T_N$ with $p$. For the parent compound ($x\,=\,0$, see Fig.\,\ref{fig:specanomaly}\,(a)), the initial suppression rate of $T_N$ is consistent with previous literature results \cite{Colombier09}. Importantly, our study goes beyond these previous studies, as it demonstrates that $T_s$ is suppressed at a lower rate than $T_N$ over the investigated $p$ range. This results in a monotonically increasing splitting $\Delta T\,=\,T_s-T_N$ as a function of $p$ from $\Delta T (p\,=\,0)\,\approx\,1\,$K  up to $\,\approx\,$3.1\,K within 2\,GPa (see Inset of Fig. \,\ref{fig:phasediagram}\,(a)). The phase diagram of the system with intermediate Co substitution level ($x\,=\,0.02$, Fig. \,\ref{fig:phasediagram}\,(b)), which is well-known to exhibit a sizable $\Delta T$ at ambient $p$, initially reveals a very similar behavior, which gives rise to an increase of $\Delta T$ with $p$. However, in this case, above $p\,\,\approx\,1.3\,$GPa, the behavior is suddenly reversed and $T_s$ is suppressed faster than $T_N$. Correspondingly, the two transitions approach each other again, which is displayed in a decreasing $\Delta T$ with $p$, and tend to merge at $p\,>\,2\,$GPa. It is interesting to note that the sudden reversal of $\Delta T$ with $p$ at $\,\approx\,1.3$\,GPa mainly results from a change of the behavior of $T_s$ with $p$, as $T_N$ shows a smooth evolution with $p$. In case of a sample with even higher Co concentration ($x\,=\,0.033$, Fig.\,\ref{fig:phasediagram}\,(c)), $T_s$ is suppressed at a higher rate than $T_N$ over the full investigated $p$ range. Thus, the initially well-separated transitions approach each other rapidly and merge at $p\,\approx\,1.5\,$GPa. We note that a recent $p$ study \cite{Ikeda18} of a sample with $x\,=\,0.025$ disclosed a monotonically decreasing $\Delta T$ with $p$ as well.

		We can summarize these observations as follows: at low $x$ and/or low $p$, $\Delta T$ increases with $p$, whereas for higher $x$ and/or higher $p$, $\Delta T$ decreases with $p$. In other words, there exists two distinct regimes in which $p$ and Co substitution either act similarly in terms of the splitting, or counteract each other, respectively. The transition between the two regimes can be induced by changing $x$, implying a critical concentration $x_c$. Alternatively, as our measurements on the $x\,=\,0.02$ sample show, application of $p$ can also result in a transition from the d($\Delta T$)/d$p\,>\,0$ to d($\Delta T$)/d$p\,<\,0$ regime at a critical pressure $p_c(x)$, as long as $x\,<\,x_c$. In terms of $x_c$, taken together with the data from Ref.\,\citen{Ikeda18}, we can infer that $0.02\,\le\,x_c\,\le\,0.025$. Previous ARPES \cite{Liu10}, as well as thermoelectric power and Hall effect measurements \cite{Mun09,Hodovanets13} at ambient $p$ revealed that there is a sudden change of the Fermi surface topology (labeled Lifshitz transition therein) as a function of $x$ in this $x_c$ range, which manifests itself particularly strongly in the afm state. Thus, this is highly suggestive of a strong correlation between the distinct $p$ response of $T_s-T_N$ and the change in the Fermi surface topology.
		
		\begin{figure}
		\includegraphics[width=\columnwidth]{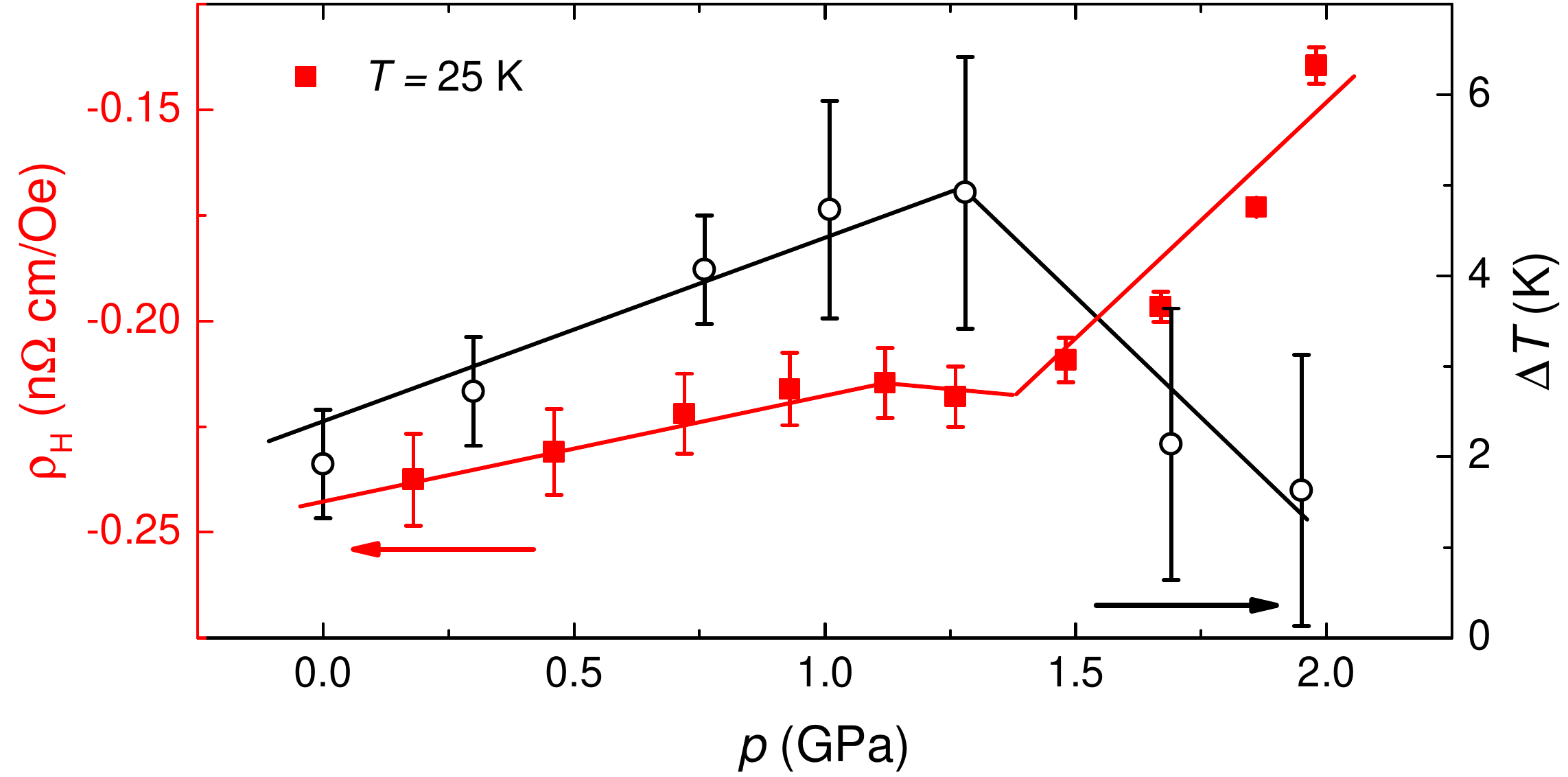} 
		\caption{Pressure-dependent Hall coefficient, $\rho_{H}$, taken at $T\,=\,$25\,K (left axis, filled squares) and $\Delta T\,=\,T_s-T_N$ (open stars) vs. $p$ for Ba(Fe$_{1-x}$Co$_x$)$_2$As$_2$ ($x\,=\,0.02$). Solid lines are guides to the eye.}
		\label{fig:Hall}
		\end{figure}
		
		\begin{figure*}
		\includegraphics[width=0.9\textwidth]{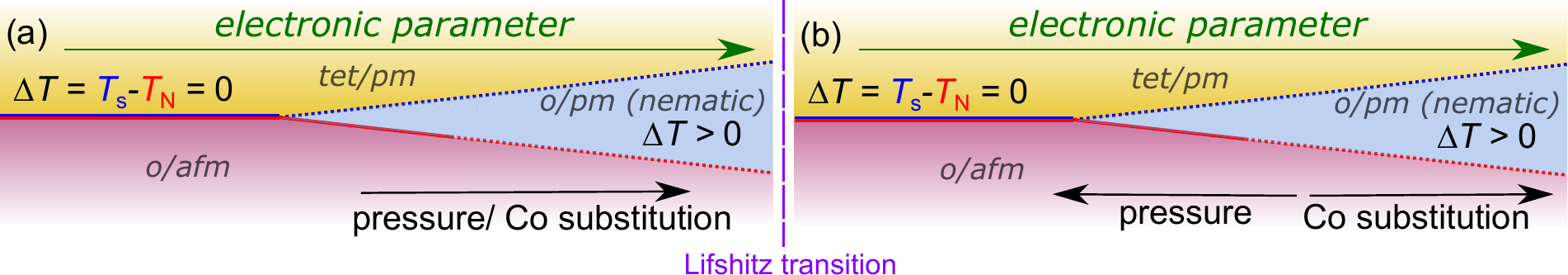} 
		\caption{Proposed schematic phase diagram of $\Delta T\,=\,T_s-T_N$ as a function of an electronic parameter, which tunes the magnetic and structural transition from a simultaneous first-order transition to well-separated second order transitions. Arrows indicate the response of Ba(Fe$_{1-x}$Co$_x$)$_2$As$_2$ to pressure and Co substitution before (a) and after (b) undergoing a change of Fermi surface topology. Solid (dotted) lines indicate first (second) order transitions, red (blue) lines refer to the antiferromagnetic (structural) phase transition.}
		\label{fig:universal}
		\end{figure*}
		
		To pinpoint this cross-correlation in the case of a $p$-induced transition between the two regimes, we examined the compound with $x\,=\,0.02$ by Hall measurements as a function of $p$ across $p_c(x\,=\,0.02)\,\approx\,1.3\,$GPa. The evolution of the Hall coefficient, $\rho_{H}$, taken at $T\,=\,25\,$K, with $p$ is presented in Fig.\,\ref{fig:Hall} (and Figs.\,\ref{fig:Hall-raw}-\ref{fig:Hall-all}), together with $\Delta T(p)$ inferred from the specific heat measurements. This data set clearly reveals a non-monotonic behavior with a kink-like change of the Hall coefficient at $p_c\,\approx\,$1.3\,GPa, at which d($\Delta T$)/d$p$ changes sign. Such a non-monotonic behavior of the Hall coefficient as a function of the clean tuning parameter $p$ on one single sample cannot result from a variation of extrinsic parameters (like disorder) and therefore strongly suggests a change in the band structure. We stress that our data sets at different temperatures (see SI Fig.\,\ref{fig:Hall-all}) reveal a feature at $p_c$ not only in the afm state, but also in the paramagnetic, tetragonal phase. Unfortunately, due to the limited availability of techniques resolving the Fermi surface under $p$, we are not able to determine the exact associated changes in the Fermi surface topology (see discussion below). Nevertheless, two important conclusions can be drawn from our experiments. First, a change in the Fermi surface topology is a generic feature of the BaFe$_2$As$_2$ phase diagram which can be induced not only by electron doping, but also by $p$ (which is not expected \textit{a priori}, as generically both tuning parameters have distinctly different effects on the band structure). This observation, together with the initial increase of splitting of $T_s$ and $T_N$, emphasizes that starting from the parent compound  Co substitution and $p$ \textit{initially} act in a very similar manner. Second, the breakdown of this analogy, which manifests itself in a different response of $T_s$ with respect to $T_N$ to tuning by $p$ and doping, is associated with a sudden change of Fermi surface topology (see Fig.\,\ref{fig:universal} for a schematic illustration).
		
		\textit{Discussion-} Our findings have important implications on the general picture of the iron-pnictide phase diagram. Even though electron doping in form of Co substitution and pressure act very similar on a gross level \cite{Kimber09,Colombier09,Colombier10b} (i.e., suppression of $T_N$ and $T_s$), our study clearly demonstrated that they are not equal tuning parameters on a finer level. 
		
		This being said, our experimental results strongly suggest that the evolution of structural order with respect to magnetic ordering in the 122 iron pnictides is governed by a more general parameter of electronic origin, as the high-$T$ paramagnetic state at $T\,\ge\,T_s$ does not undergo any structural change \cite{Ni08,Kimber09} as a function of $x$ or $p$ in this $x$ and $p$ range. This conclusion is therefore consistent with any model of electronic nematicity. Moreover, this parameter is likely related to detailed Fermi surface topology. As a consequence, sudden changes in the Fermi surface topology result in a non-monotonic evolution of this parameter as a function of experimentally accessible tuning parameters. 
		
		Indeed, the microscopic model of Refs.\,\citen{Fernandes12,Fernandes14} mapped the phase diagram, evolving from a simultaneous first-order magneto-structural transition to well-separated second-order transitions (see Fig.\,\ref{fig:universal}), onto a single parameter $\alpha$. In this strongly simplified two-dimensional model, $\alpha$ is mainly affected by changes in the chemical potential, $\mu$, as well as the ellipticity of the electron pockets, $\delta m$, and therefore represents a parameter, which characterizes the nesting conditions of the Fermi surface. If such a single parameter indeed exists, it is required for fundamental reasons \cite{Fernandes12}, based on symmetry arguments, that the merged magneto-structural transition is first order in character, although potentially only very weak. Even though this makes a definite experimental proof extremely difficult in the presence of disorder and/or small $p$ inhomogeneities, our results on the two border compounds of this study reveal a distinctly different behavior (see SI, Fig.\,\ref{fig:broadening}): whereas in the pure compound the application of $p$ leads to an enhanced broadening of the afm peak, the peak width is almost unaffected, if not even slightly reduced, in the case of $x\,=\,0.033$. This observation suggests that $p$ modifies the character of the transition at $x\,<\,x_c$ and $x\,>\,x_c$ in a very different manner and can therefore be considered as an indication that the merged transition is indeed rather first order in character. 
		
		It is also worthwhile to note that composite phase diagrams as a function of Co substitution and $p$ (see SI Fig.\,\ref{fig:unified}) at $x\,<\,x_c$ and $p\,<\,p_c$ demonstrate that $T_s$ is almost similar for a given $T_N(x)\,=\,T_N(p)$. This does not only point towards a similar origin of $T_s$ and $T_N$, but also strengthens the viewpoint of a parameter of electronic origin governing the evolution of nematic order with respect to magnetic order in the phase diagram.
		
		In a next step, it is important to identify how small $p$ and small amount of Co substitutions affect the three-dimensional Fermi surface at and beyond the sudden change of Fermi surface topology at $x_c$ or $p_c$. A preliminary attempt to describe these Fermi surface changes by standard DFT band-structure calculations \cite{Borisov19} called for extended approaches, which capture the presence of sizable spin fluctuations. This study is potentially feasible, but goes beyond the present study. If successful, these results will then form the basis to verify different microscopic models of electronically-driven nematicity in the 122 family iron pnictides.

		Last, we want to point out which implications our results might have for the emergence of superconductivity in the BaFe$_2$As$_2$ phase diagram. Previous studies \cite{Ni08,Canfield09,Avci12,Luetkens09,Rotundu09,Canfield10,Sefat08,Takahashi08,Rotter08,Alireza09} have established that $p$ as well as Co substitution induce superconductivity once $T_s$ and $T_N$ are sufficiently suppressed. It was proposed that superconducting pairing in this series is enhanced by either spin or nematic fluctuations \cite{Chu12,Kuo16} which originate from the respective, putative quantum-critical points \cite{Zhou13,Lederer15,Analytis14}. It is therefore noteworthy that the application of $p$ on samples with $x\,>\,x_c$ (see Ref.\,\citen{Colombier10b} and SI Fig.\,\ref{fig:sc}) induces superconductivity, although $T_s$ and $T_N$ merge and likely become a weak first-order transition. Consequently, if critical fluctuations associated with a magnetic and/or nematic quantum-critical point promote superconductivity, it is crucial to identify the impact of fluctuations on superconductivity in the presence of a weak first-order transition. 
	
\textit{Conclusion-} We performed the first systematic study of the structural and magnetic transition temperatures $T_s$ and $T_N$ under pressure in Ba(Fe$_{1-x}$Co$_x$)$_2$As$_2$. Our results demonstrate that the pressure response of $T_s$, compared to $T_N$, is strongly modified when the system suddenly changes its Fermi surface topology either as a function of Co substitution or pressure. We argue that this observation speaks in favor of a Fermi-surface characterizing parameter of electronic origin that governs the evolution of nematic order in the iron-pnictide phase diagram. This result therefore represents an important experimental benchmark, with clear critical pressures and concentrations, for understanding the origin of nematicity and its relation to superconductivity in the iron-pnictide family.

\begin{acknowledgments}
\textit{Acknowledgments- }We thank V. Borisov and R. Valent\'{i} for their heroic efforts in preliminary DFT calculations, A. Kreyssig, R. Fernandes and A. Kaminski for useful discussions, N. Ni and A. Thaler for growing the Ba(Fe$_{1-x}$Co$_x$)$_2$As$_2$ crystals used in the study and G. Drachuck for assistance with the specific heat measurements. Work at the Ames Laboratory was supported by the U.S. Department of Energy, Office of Science, Basic Energy Sciences, Materials Sciences and Engineering Division. The Ames Laboratory is operated for the U.S. Department of Energy by Iowa State University under Contract No. DEAC02-07CH11358. E.G. and L.X. were funded, in part, by the Gordon and Betty Moore Foundation’s EPiQS Initiative through Grant GBMF4411. L.X. was funded, in part, by the W. M. Keck Foundation.
\end{acknowledgments}

\appendix
	\section{Specific heat under pressure}
	
	\subsection{Detailed specific heat data}
	
\begin{center}
		\begin{figure*}[h!]
		\includegraphics[width=0.9\textwidth]{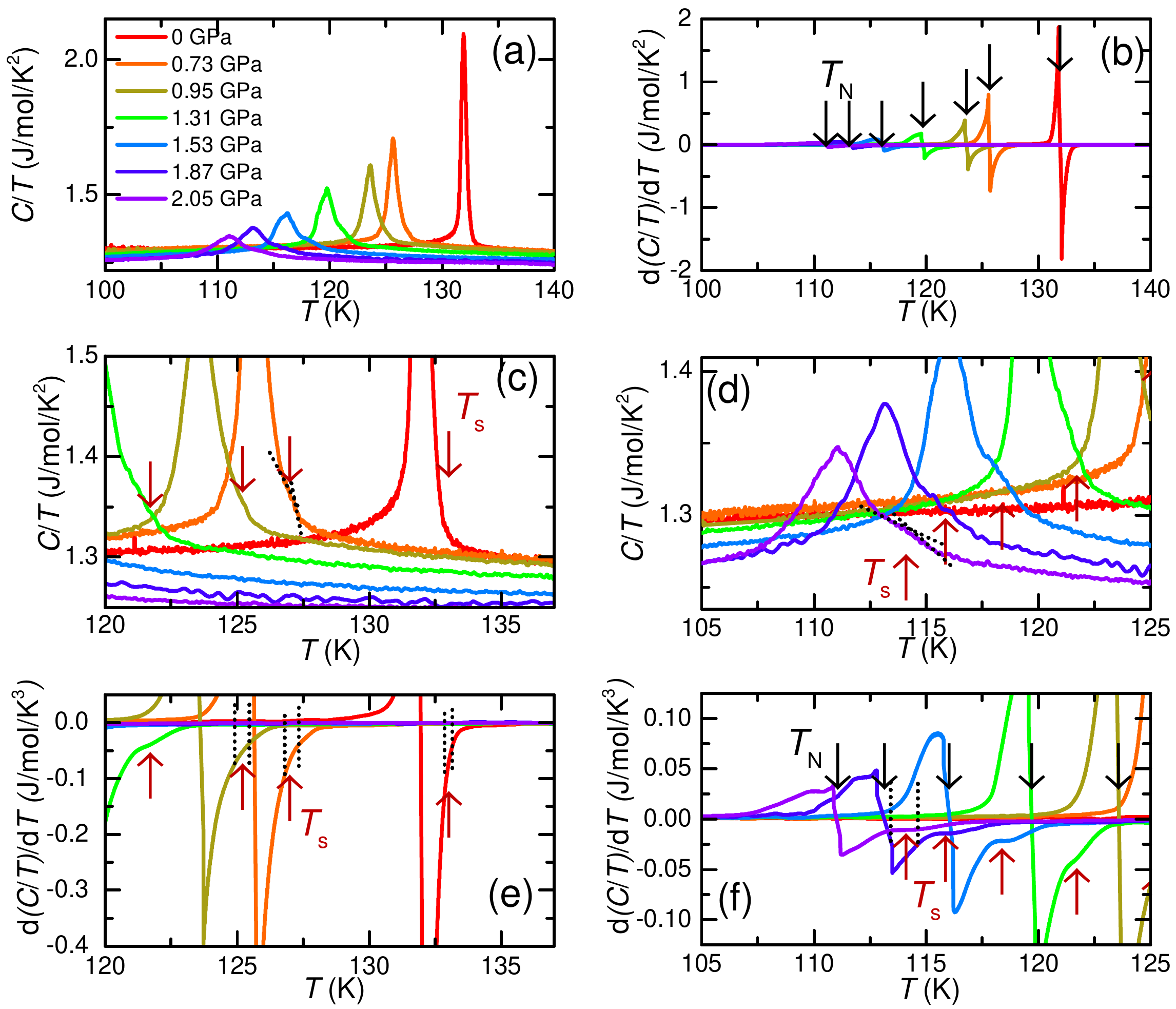} 
		\caption{(a) Specific heat divided by temperature, $C/T$, vs. temperature $T$ of BaFe$_2$As$_2$ at different pressures, $p$, between 0\,GPa and 2.05 GPa; (b) Derivative of $C/T$ with respect to $T$, d$(C/T)$/d$T$, for the same pressure values as depicted in (a); (c) and (d) Blow-ups of the data, presented in (a); (e) and (f) Blow-ups of the data, presented in (b). Arrows and lines indicate the criteria to determine the antiferromagnetic and structural transition temperatures at $T_{N}$ and $T_{S}$, respectively.}
		\label{fig:Ba122}
		\end{figure*}
		\end{center}

\begin{figure*}[h!]
		\begin{center}
		\includegraphics[width=0.9\textwidth]{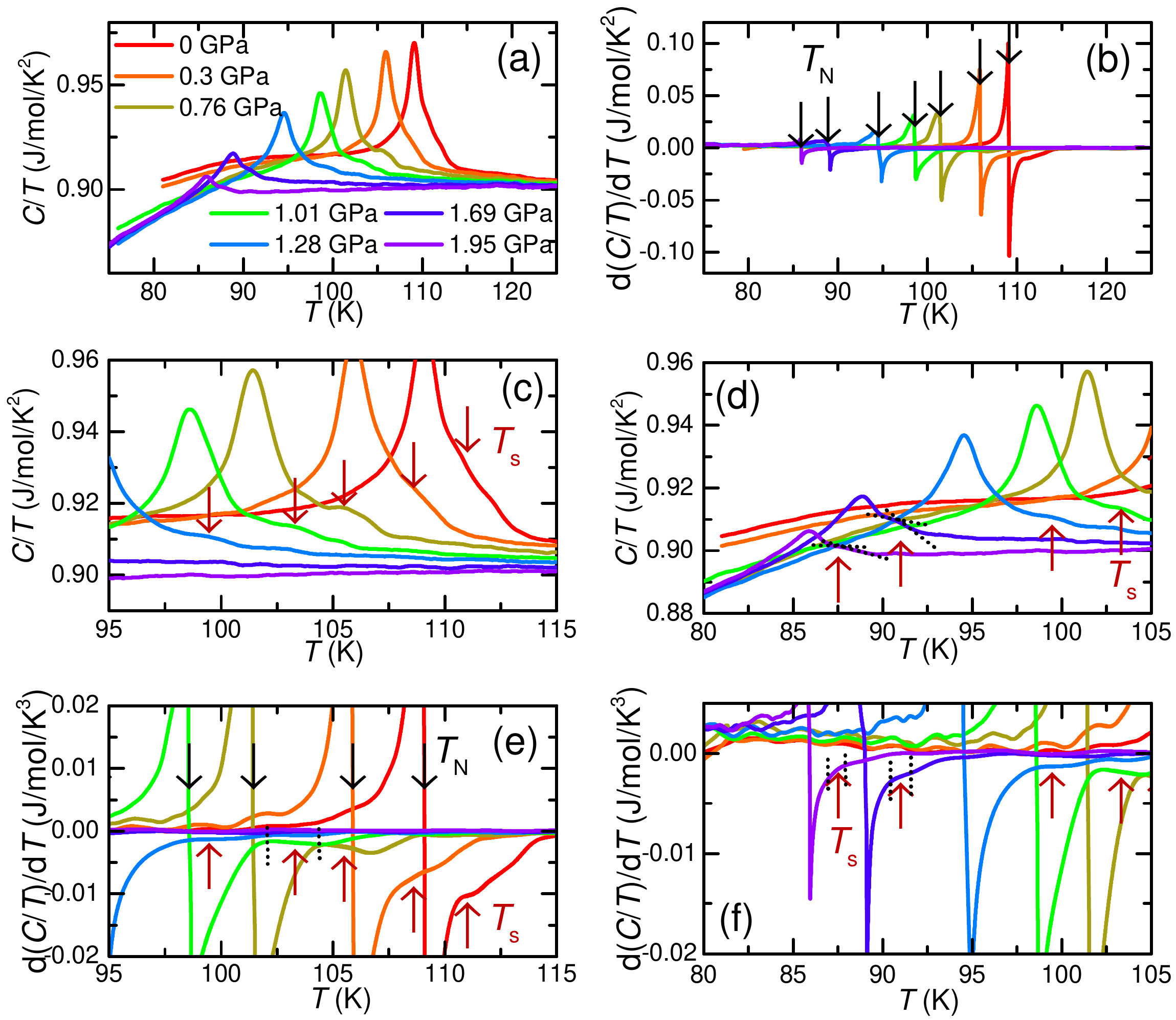} 
		\caption{(a) Specific heat divided by temperature, $C/T$, vs. temperature $T$ of Ba(Fe$_{0.98}$Co$_{0.02}$)$_2$As$_2$ at different pressures, $p$, between 0\,GPa and 1.95 GPa. Data were slightly offset with respect to each other for clarity; (b) Derivative of $C/T$ with respect to $T$, d$(C/T)$/d$T$, for the same pressure values as depicted in (a); (c) and (d) Blow-ups of the data, presented in (a); (e) and (f) Blow-ups of the data, presented in (b). Arrows and lines indicate the criteria to determine the antiferromagnetic and structural transition temperatures at $T_{N}$ and $T_{S}$, respectively.}
		\label{fig:BaCo122}
		\end{center}
		\end{figure*}

\begin{figure}[h!]
		\begin{center}
		\includegraphics[width=\columnwidth]{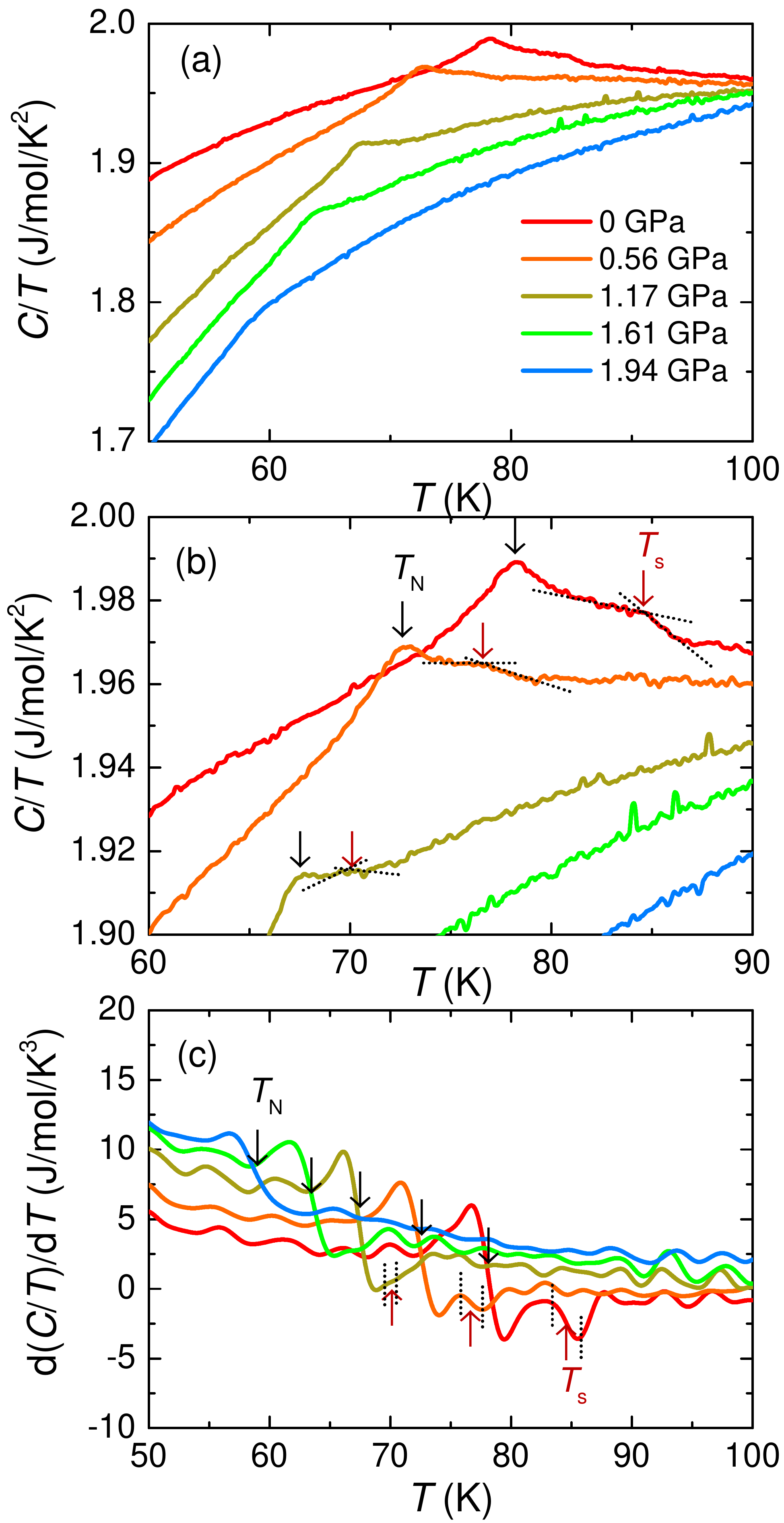} 
		\caption{(a) Specific heat divided by temperature, $C/T$, vs. temperature $T$ of Ba(Fe$_{0.967}$Co$_{0.033}$)$_2$As$_2$ at different pressures, $p$, between 0\,GPa and 1.95 GPa. Data were slightly offset with respect to each other for clarity; (b) Blow-up of the data, presented in (a); (c) Derivative of $C/T$ with respect to $T$, d$(C/T)$/d$T$, for the same pressure values as depicted in (a,b). Arrows and lines indicate the criteria to determine the antiferromagnetic and structural transition temperatures at $T_{N}$ and $T_{S}$, respectively.}
		\label{fig:BaCo122-2}
		\end{center}
		\end{figure}

	Figures \ref{fig:Ba122}-\ref{fig:BaCo122-2} show the raw specific heat data (Figs.\,\ref{fig:Ba122} (a,c,d), \ref{fig:BaCo122} (a,c,d) and \ref{fig:BaCo122-2} (a,b)) of the three Ba(Fe$_{1-x}$Co$_x$)$_2$As$_2$ samples with $x\,=\,0, 0.02$ and 0.033, studied in this work, as well as the derivative of these data sets (Figs.\,\ref{fig:Ba122} (b,e,f), \ref{fig:BaCo122} (b,e,f) and \ref{fig:BaCo122-2} (c)). The data shown in the main manuscript in Fig.\,1 were obtained from the raw data sets presented in Figs. \ref{fig:Ba122}-\ref{fig:BaCo122-2} by subtracting a smooth background contribution. The latter was obtained by fitting the specific heat data for $T_N-15\,$K$\,\le\,T\,\le\,T_N-5\,$K and $T_s+5\,$K$\,\le\,T\,\le\,T_s+15\,$K with a polynomial function of the order of 3.
	
\begin{figure*}[h!]
		\begin{center}
		\includegraphics[width=0.6\textwidth]{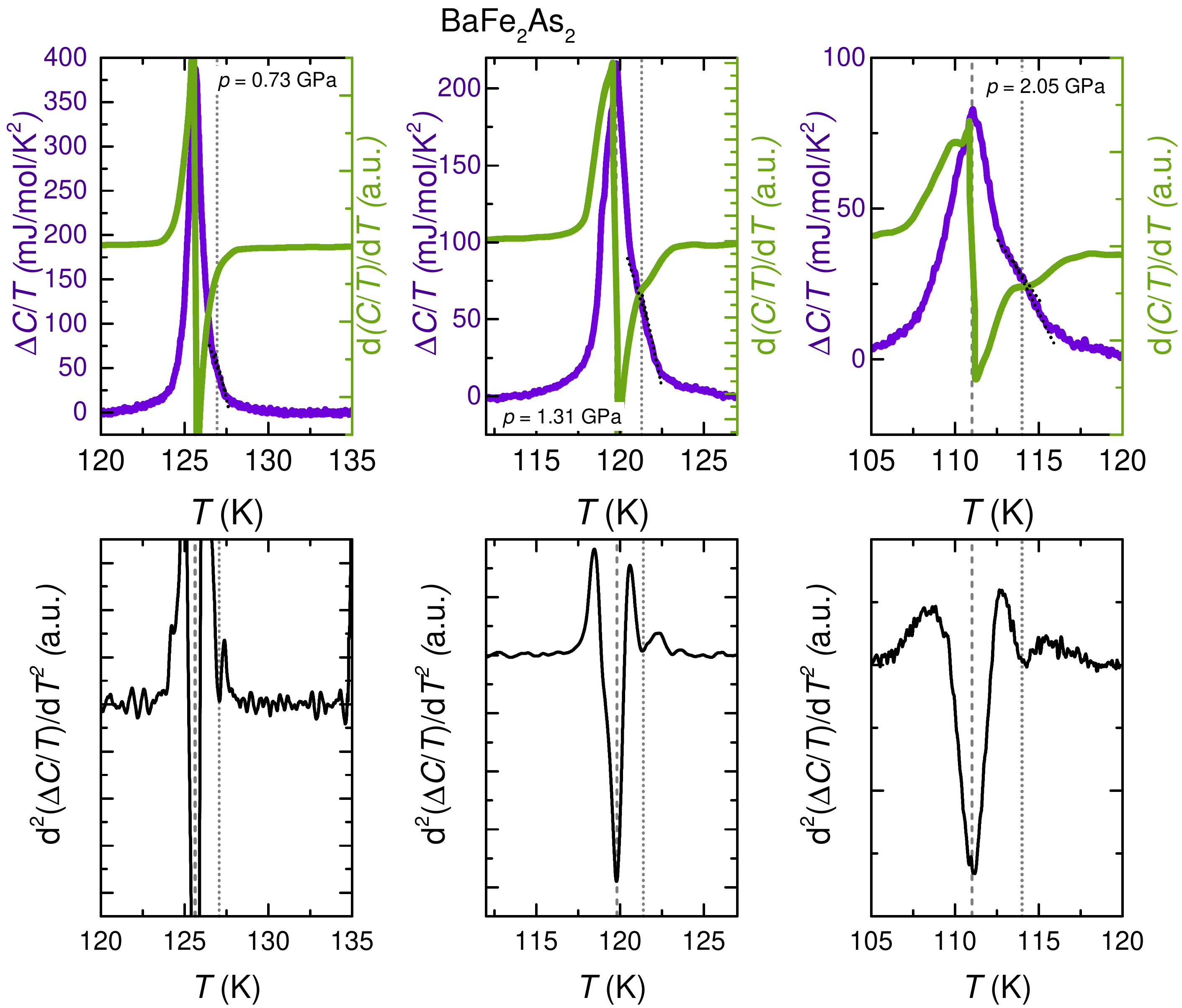} 
		\caption{Comparison of specific heat data, $\Delta C/T$ (top panel, left axis), first derivative, d$(C/T)/$d$T$ (top panel, right axis) and second derivative d$^2 (\Delta C/T)/$d$T^2$ (bottom panel) for three selected pressures of BaFe$_2$As$_2$.}
		\label{fig:criteriondiscussion1}
		\end{center}
		\end{figure*}

\begin{figure*}[h!]
		\begin{center}
		\includegraphics[width=\textwidth]{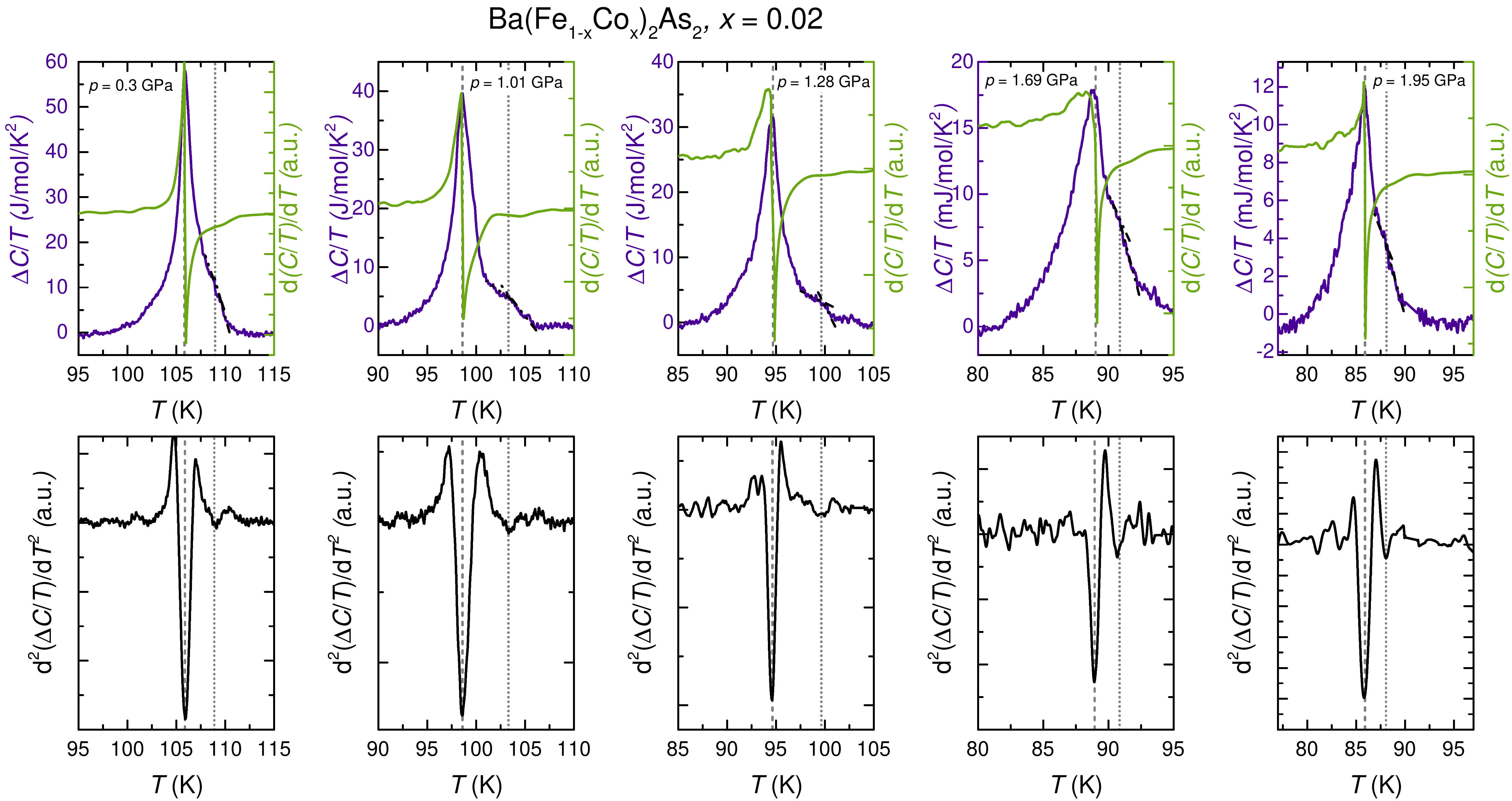} 
		\caption{Comparison of specific heat data, $\Delta C/T$ (top panel, left axis), first derivative, d$(C/T)/$d$T$ (top panel, right axis) and second derivative d$^2 (\Delta C/T)/$d$T^2$ (bottom panel) for five selected pressures of Ba(Fe$_{1-x}$Co$_x$)$_2$As$_2$ with $x\,=\,$0.02.}
		\label{fig:criteriondiscussion2}
		\end{center}
		\end{figure*}

\begin{figure*}[h!]
		\begin{center}
		\includegraphics[width=0.6\textwidth]{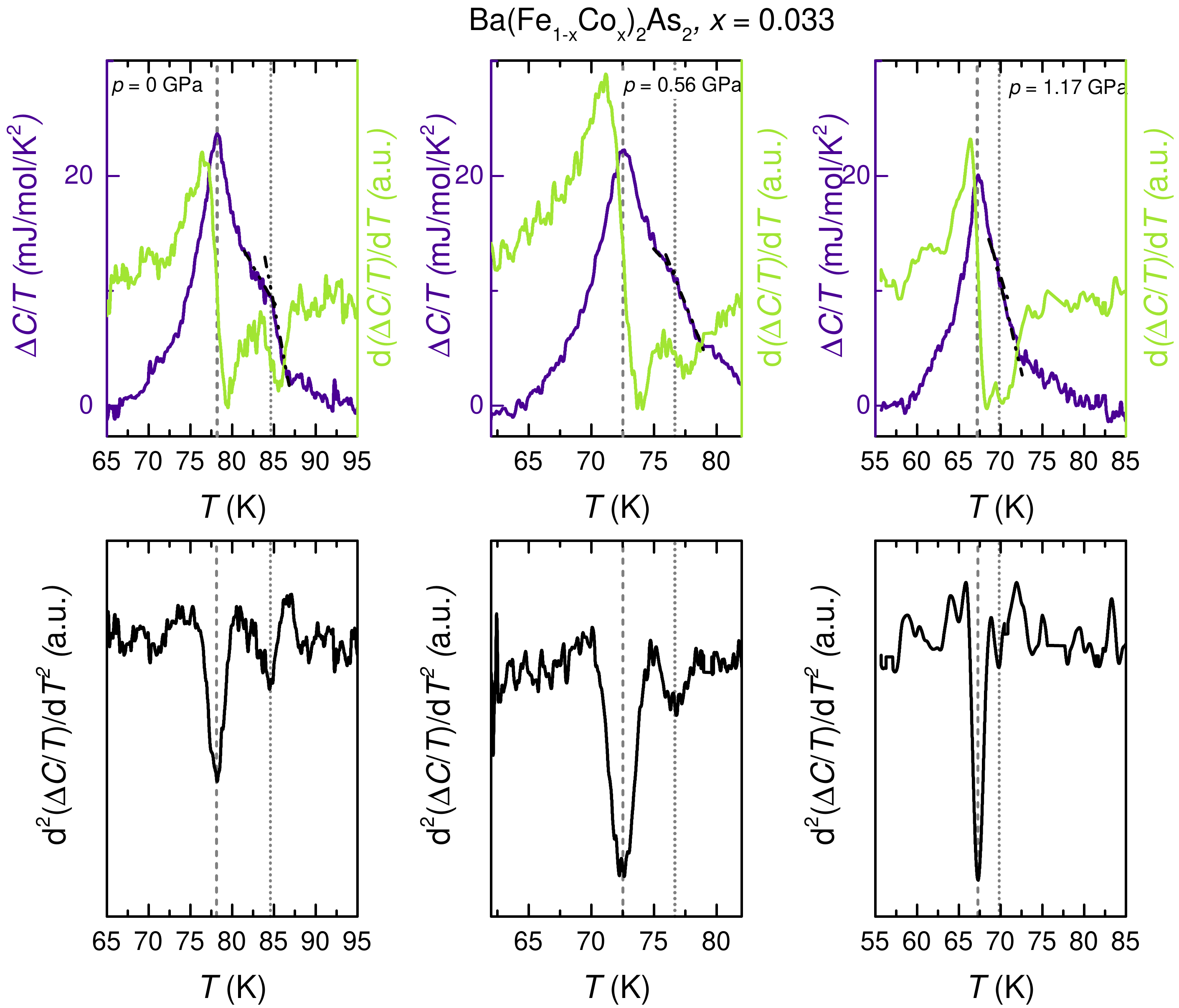} 
		\caption{Comparison of specific heat data, $\Delta C/T$ (top panel, left axis), first derivative, d$(\Delta C/T)/$d$T$ (top panel, right axis) and second derivative d$^2 (\Delta C/T)/$d$T^2$ (bottom panel) for three selected pressures of Ba(Fe$_{1-x}$Co$_x$)$_2$As$_2$ with $x\,=\,$0.03.}
		\label{fig:criteriondiscussion3}
		\end{center}
		\end{figure*}

		These data sets, depicted in Figs. \ref{fig:Ba122}-\ref{fig:BaCo122-2}, are used to determine the transition temperatures $T_N$ and $T_s$. The criteria used are schematically illustrated in each panel and will be discussed in the following. The larger peak in $C/T$ is associated with the transition into the antiferromagnetic state at $T_N$. Correspondingly, all data sets reveal a sharp step in d$(C/T)$/d$T$ (see Figs.\,\ref{fig:Ba122}(b), Fig.\,\ref{fig:BaCo122}\,(b) and Fig.\,\ref{fig:BaCo122-2} (c)). The midpoint of this step-like feature is used to infer $T_N$. Equally, one can refer to the resulting minimum in the second derivative of the specific heat data, which are exemplarily shown for each compound in \ref{fig:criteriondiscussion1}-\ref{fig:criteriondiscussion3}. As shown in each individual panel by grey dashed lines, the so-derived $T_N$ values correspond to the peak position of the $\Delta C/T$ data, which are marked by arrows in Fig.\,1 of the main manuscript.

		If the two specific heat peaks, associated with the magnetic and structural transition, are well-enough separated, then the specific heat peak at $T_s$ will also result in a step-like feature in the first derivative, albeit smaller in size. For example, such a step-like feature is shown on an enlarged scale in figure \ref{fig:criteriondiscussion3} below for $x$\,=\,0.033 and $p\,=\,0.56\,$GPa. The midpoint of this step-like feature can be used to infer $T_s$. Again, the so-inferred $T_s$ values correspond well to the position of the kink in $\Delta C/T$ (schematically indicated by the intersection of two straight lines in Figs.\,\ref{fig:Ba122}(c),(d), Figs.\,\ref{fig:BaCo122} (c),(d) and Figs.\,\ref{fig:BaCo122-2}(b)) and the minimum in the second derivative, as visualized by the grey dotted line in Figs. \ref{fig:criteriondiscussion1}-\ref{fig:criteriondiscussion3}. 
In case $T_s$ and $T_N$ are closer, the superposition of the two specific heat peaks obviously manifests itself also in the first derivative. As a consequence, the first derivative depicts a plateau rather than a sharp step-like feature at $T_s$ (see for example figure \ref{fig:criteriondiscussion1} below, for $x\,=\,0$ and $p\,=\,2.05$ GPa and Figs.\,\ref{fig:Ba122}\,(e),(f), Figs.\,\ref{fig:BaCo122}\,(e),(f) and Fig.\,\ref{fig:BaCo122-2}\,(c)). The midpoint of this plateau can be used to infer $T_s$. In fact, a clear fingerprint of this plateau can be observed when considering the second derivative, which shows a shallow minimum exactly at the midpoint of the plateau in the first derivative and the kink in $\Delta C/T$ (see grey dotted lines in Fig.\,\ref{fig:criteriondiscussion1}-\ref{fig:criteriondiscussion3} for comparison of data, first and second derivative; see intersection of dotted lines in Figs.\,\ref{fig:Ba122}-\ref{fig:BaCo122-2} which illustrate the kink position in $C/T$ for various data points as well as the corresponding plateau features in d$(C/T)$/d$T$).

		For computing the derivatives of our specific heat data, the raw data were smoothened. In doing so, care was taken that smoothening does not result in a significant shift of the anomalies in temperature. This typically resulted in a sliding average over a temperature window of less than 0.3\,K (while raw data spacing is less than 1\,mK). From the good agreement of the peak position at $T_N$ or the kink position in $T_s$ in $\Delta C/T$ with the step-like features in the first derivative as well as the minima in the second derivative (see dashed and dotted lines in Figs.\,\ref{fig:criteriondiscussion1}-\ref{fig:criteriondiscussion3}), we can conclude that the error due to smoothening is comparably small. Nevertheless, whereas the determination of $T_N$ results only in a small error due to the sharpness of the features in d($\Delta C/T$)/d$T$ and d$^2(\Delta C/T$)/d$T^2$, there is certainly a larger error bar involved in the determination of $T_s$. We estimate this error from half of the width of the step-like features (or plateau-like features) in the first derivative, d($\Delta C/T$)/d$T$ and crosschecked these with the full width at half maximum of the minima in the second derivative. The resulting errors are of symbol size in the main panels of Fig.\,2 of the main manuscript. These errors directly result in error bars in the inferred $\Delta T$ values which are clearly depicted in the insets of Fig.\,2 of the main manuscript.
		
		\subsection{First vs. second-order transition}
		
\begin{figure}[h!]
		\begin{center}
		\includegraphics[width=\columnwidth]{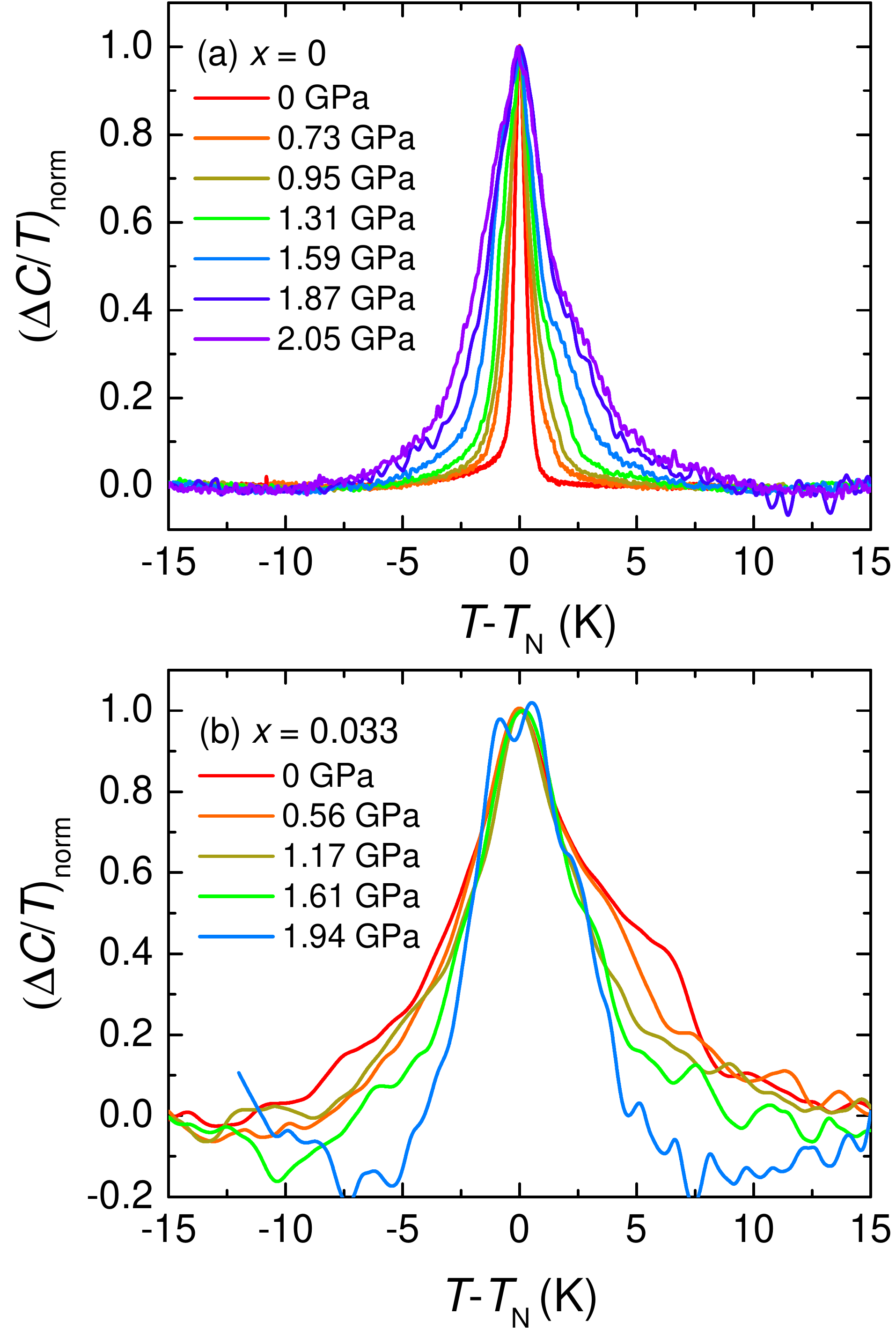} 
		\caption{Anomalous specific heat contribution, $\Delta C/T$, at different pressures, each normalized to the maximum value at $T_N$ vs. $T-T_N$ for Ba(Fe$_{1-x}$Co$_x$)$_2$As$_2$ with $x\,=\,0$ (a) and 0.033 (b).}
		\label{fig:broadening}
		\end{center}
		\end{figure}

		As indicated in the main text, from a symmetry point of view, it is required that the merged magneto-structural transition is a first-order transition, if it is smoothly connected to two separated second-order transitions. At the same time, this implies that the first-order transition close to the merging point is probably rather weak. This, together with experimental uncertainties resulting from disorder and/or small pressure inhomogeneities, make a definite experimental distinction between first- and second-order transitions extremely difficult. To investigate potential changes in the character of the phase transition, we focus here on an analysis of the specific heat peak, which is associated with the afm ordering, of the $x\,=\,0$ and 0.033 compounds. To this end, in Fig.\,\ref{fig:broadening}, we normalized the data shown in Fig.\,1 of the main manuscript to their respective peak temperature and peak specific heat value. For $x\,=\,0$, we find a monotonically increasing peak width as a function of pressure across the full pressure range. In contrast, the width of the specific heat peak for $x\,=\,0.033$ is almost unaffected by changing pressure, if not even a bit reduced. In the main text, we demonstrated that pressure on the $x\,=\,0$ on the one hand results in an increased splitting of $T_s$ and $T_N$ and proposed that this moves the system further towards the limit of two well-separated second-order transitions. Thus, the increase in peak width is fully consistent with this proposal. On the other hand, we showed that pressure on the $x\,=\,0.033$ compound causes both transitions to merge at $p\,\approx\,1.5\,$GPa, which has to result in a change from second order to first order. The markedly different evolution of the peak width of the compound with $x\,=\,0.033$, compared to the $x\,=\,0$ compound, can be considered as an indication that the character of the transition changes from second order to first order upon applying pressure.
		
		\subsection{Unified phase diagrams}
		
\begin{figure*}[h!]
		\begin{center}
		\includegraphics[width=0.9\textwidth]{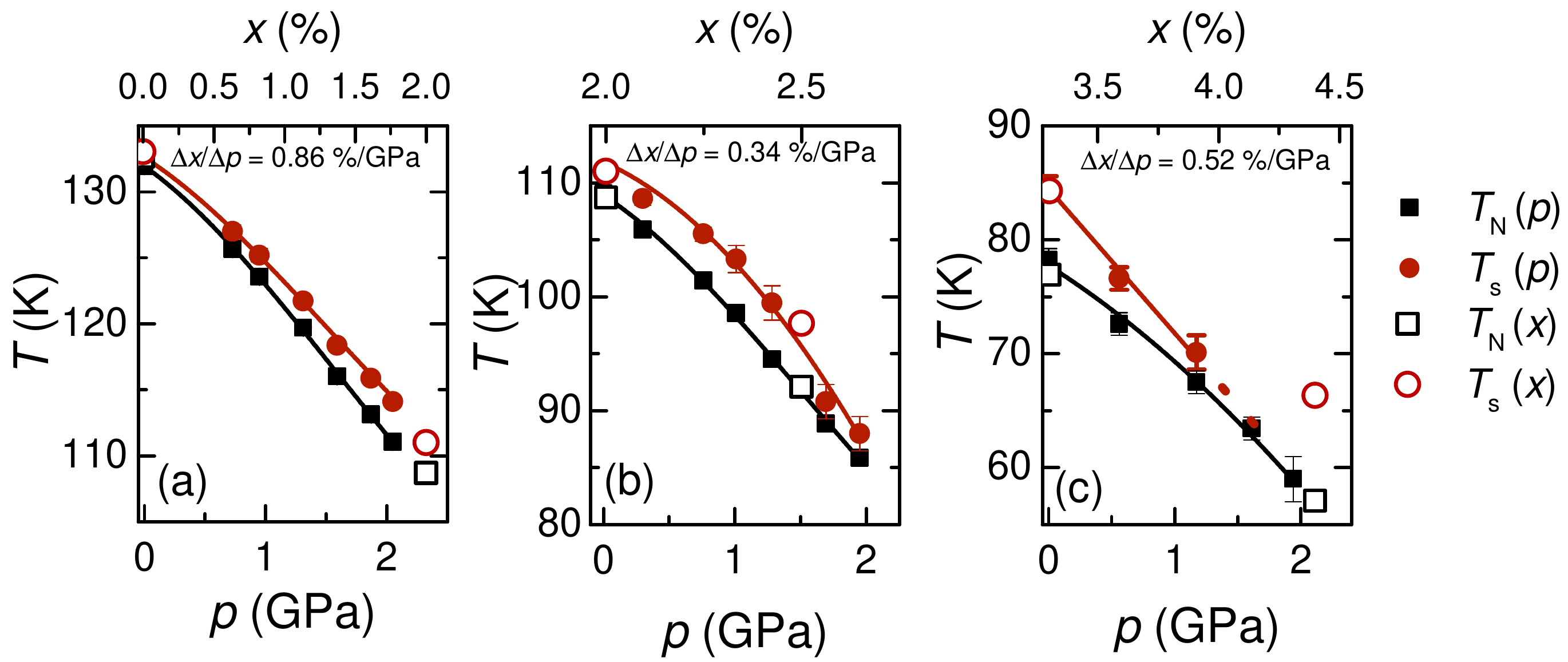} 
		\caption{Unified phase diagrams as a function of pressure, $p$, (bottom axis) and Co substitution $x$ (top axis) for Ba(Fe$_{1-x}$Co$_x$)$_2$As$_2$ ($x\,=\,0$ (a), 0.02 (b) and 0.033 (c)). Full (open) black squares correspond to $T_N$ as a function of $p$ ($x$), full (open) brown circles correspond to $T_s$ as a function of $p$ ($x$). Data as a function of $x$ was reproduced from literature \cite{Ni08}.}
		\label{fig:unified}
		\end{center}
		\end{figure*}

		The present study focuses on an investigation of the relation of antiferromagnetic and structural transition temperatures $T_N$ and $T_s$ in BaFe$_2$As$_2$ as a function of pressure and Co substitution. Our main finding is that there exists two distinctly different regimes, in which application of pressure results either in an increase or a decrease of the splitting of $T_s$ and $T_N$. In comparison to the effect of Co substitution, this implies that in the first regime Co substitution and pressure act in a similar manner, whereas in the latter regime Co substitution and pressure counteract. To further quantify this statement, we present in Fig.\,\ref{fig:unified} unified phase diagrams, in which we compose the phase diagrams as a function of $p$, determined in the present work, with those as a function of $x$, reproduced from Ref.\,\citen{Ni08}.
		
		To construct these composite phase diagrams, we scaled each $T$-$p$ data set for a given $x$, in such a way, that the $T_N(p)$ values match the $T_N(x)$ values of the $T$-$x$ phase diagram. Remarkably, this procedure reveals different values for the scaling parameters $\Delta x/\Delta p$ starting from (0.0086$\pm$0.0015)/GPa for $x\,=\,0$ to (0.0034$\pm$0.0015)/GPa for $x\,=\,0.02$ to (0.00052$\pm$0.0015)/GPa for $x\,=\,0.033$. It seems likely that this behavior is related to the electronic Lifshitz transition which occurs in the $x\,=\,0.02$ sample at $p_c\,\approx\,1.3\,$GPa or as a function of $x$ at $x_c\,\approx\,0.022$.
		
		First, we focus on the evolution of $T_s$ in these unified phase diagrams. Despite the comparably low data density as a function of $x$, it can be clearly seen that in the case of the samples with $x\,=\,0$ and the sample with $x\,=\,0.02$ below the critical pressure $p_c\,\approx\,1.3\,$GPa (Fig.\,\ref{fig:unified}\,(a) and (b)), the behavior as a function of $x$ and $p$ are not only qualitatively similar, but also in first approximation on a quantitative level: $T_s$ is almost identical for a given $T_N$. Obviously, this analogy breaks down above $p_c$ for the $x\,=\,0.02$ sample and, in particular, for the $x\,=\,$0.033 (Fig.\,\ref{fig:unified}\,(b) and (c)), as in this regime Co substitution and pressure counteract. 
		
		The observation of an almost perfect quantitative matching of $T_s$ and $T_N$ for low Co substitution and/or low pressure is remarkable. Taken together with the fact that pressure as well as Co substitution independently induce an electronic Lifshitz transition (see main text) at a similar $T_N$, this strengthens the present result that initially (i.e., before the system undergoes a sudden change of Fermi surface topology) Co substitution and pressure act similarly. When thinking in terms of a single parameter, which governs the evolution of nematic order with respect to magnetic order in BaFe$_2$As$_2$, the unified phase diagram suggests that this parameter is then initially modified in a similar manner by Co substitution and pressure.
		
		\subsection{Relation to superconductivity}
		
\begin{figure}[h!]
		\begin{center}
		\includegraphics[width=\columnwidth]{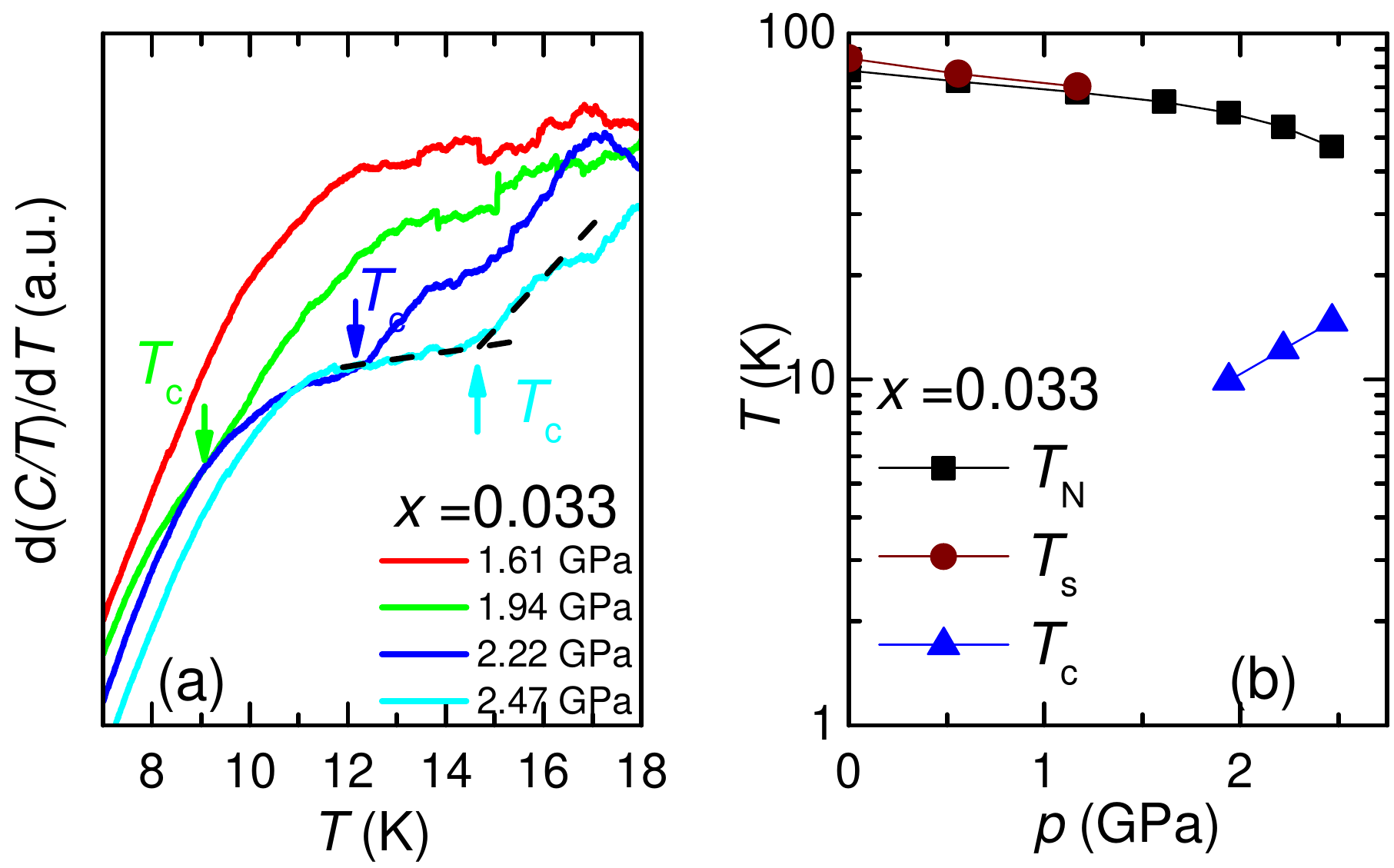} 
		\caption{(a) Derivative of low-temperature specific heat data on Ba(Fe$_{1-x}$Co$_x$)$_2$As$_2$ ($x\,=\,0.033$) at selected pressures between 1.61\,GPa to 2.47\,GPa. The kink in the data can be assigned to the superconducting phase transition $T_c$; (b) $T$-$p$ phase diagram of Ba(Fe$_{1-x}$Co$_x$)$_2$As$_2$ ($x\,=\,0.033$). Black squares correspond to the antiferromagnetic transition at $T_N$, brown circles to the structural transition at $T_s$ and blue triangles to the superconducting transition at $T_c$. The evolution of $T_c$ at lower pressures cannot be resolved in our experiment due to limitations in the lowest accessible temperature as well limitations in the resolution of broadened features.}
		\label{fig:sc}
		\end{center}
		\end{figure}

		The search for pressure-induced superconductivity in Ba(Fe$_{1-x}$Co$_x$)$_2$As$_2$ by specific heat measurements is somewhat limited by the lowest accessible temperature in these experiments ($T\,>\,5\,$K) and the highest (reliably) achievable pressure $p\,\approx\,2.3\,$GPa. In Fig.\,\ref{fig:sc}\,(a), we show low-temperature specific heat data on the sample with highest concentration in this study ($x\,=\,0.033$) at highest pressures. The data is presented in a d$(C/T)$/d$T$ vs. $T$ representation to better visualize the salient feature associated with the superconducting transition at $T_c$. We identify the kink in d$(C/T)$/d$T$, which corresponds to a broad step-like feature in $C/T$ (as well known from ambient-pressure studies on the underdoped Ba(Fe$_{1-x}$Co$_x$)$_2$As$_2$ series \cite{Ni08}), as the signature of the superconducting phase transition. This feature can clearly be resolved at $p\,\ge\,1.94$\,GPa and moves to higher temperatures with increasing $p$ (see Fig.\,\ref{fig:sc}\,(b)). This data set therefore supports the emergence of superconductivity under pressure in a situation in which the magnetic and structural transitions are merged into one, which is likely (even if weakly) first order in character.
	
		The present data set was collected on a sample, which did not show any signature of superconductivity for $T\,\ge\,2\,$K at ambient pressure. As a consequence, it is difficult to infer an onset pressure for superconductivity. Nonetheless, it is interesting to note that a clear signature of superconductivity with $T_c\,\approx\,9\,$K is detected in this sample, once $T_N$ is suppressed to $\approx\,$60\,K. In previous pressure experiments on the parent compound BaFe$_2$As$_2$ \cite{Colombier09}, zero resistance below $T_c\,\approx\,$10\,K was also detected when $T_N$ was sufficiently suppressed to $\approx\,60\,$K (at $p\,\approx\,$4\,GPa). This comparison might highlight that the suppression of $T_N$ is crucial for superconductivity to develop.
	
	\section{Hall effect under pressure}
	
	\subsection{Experimental Details}
	
	For measurements of the Hall coefficient, a sample of Ba(Fe$_{1-x}$Co$_x$)$_2$As$_2$ with $x\,=\,0.02$ was cut and cleaved into a plate-like crystal with dimensions $1\,\times\,0.84\,\times\,0.033\,$mm$^3$. Current and voltage contacts were carefully attached using Epo-tek H20E silver epoxy. Current contacts were applied to cover the two opposite ends of the crystal to ensure as uniform of a current density as possible. Voltage contacts were applied to the two remaining side faces of the crystal. Data was collected using the ACT option of the Quantum Design PPMS (Physical Property Measurement System). Polarity of the magnetic field was switched to subtract any magnetoresistive component due to a small misalignment of voltage contacts. The Hall resistivity $\rho_{xy}$ was therefore calculated as an odd-in-field component via $\rho_{xy}\,=\,(\rho_{+} - \rho_{-})/2$ with $\rho_{+}$ and $\rho_{-}$ being the resistance in positive and negative magnetic field, respectively. Pressure was created in a piston-cylinder pressure cell made out of CuBe/Ni-Cr-Al. A 4:6 mixture of light mineral oil and n-pentane was used as a pressure transmitting medium. The given pressure values correspond to the ones determined at low temperatures via the shift of the critical temperature of elemental lead.

		\begin{figure}[h!]
		\begin{center}
		\includegraphics[width=\columnwidth]{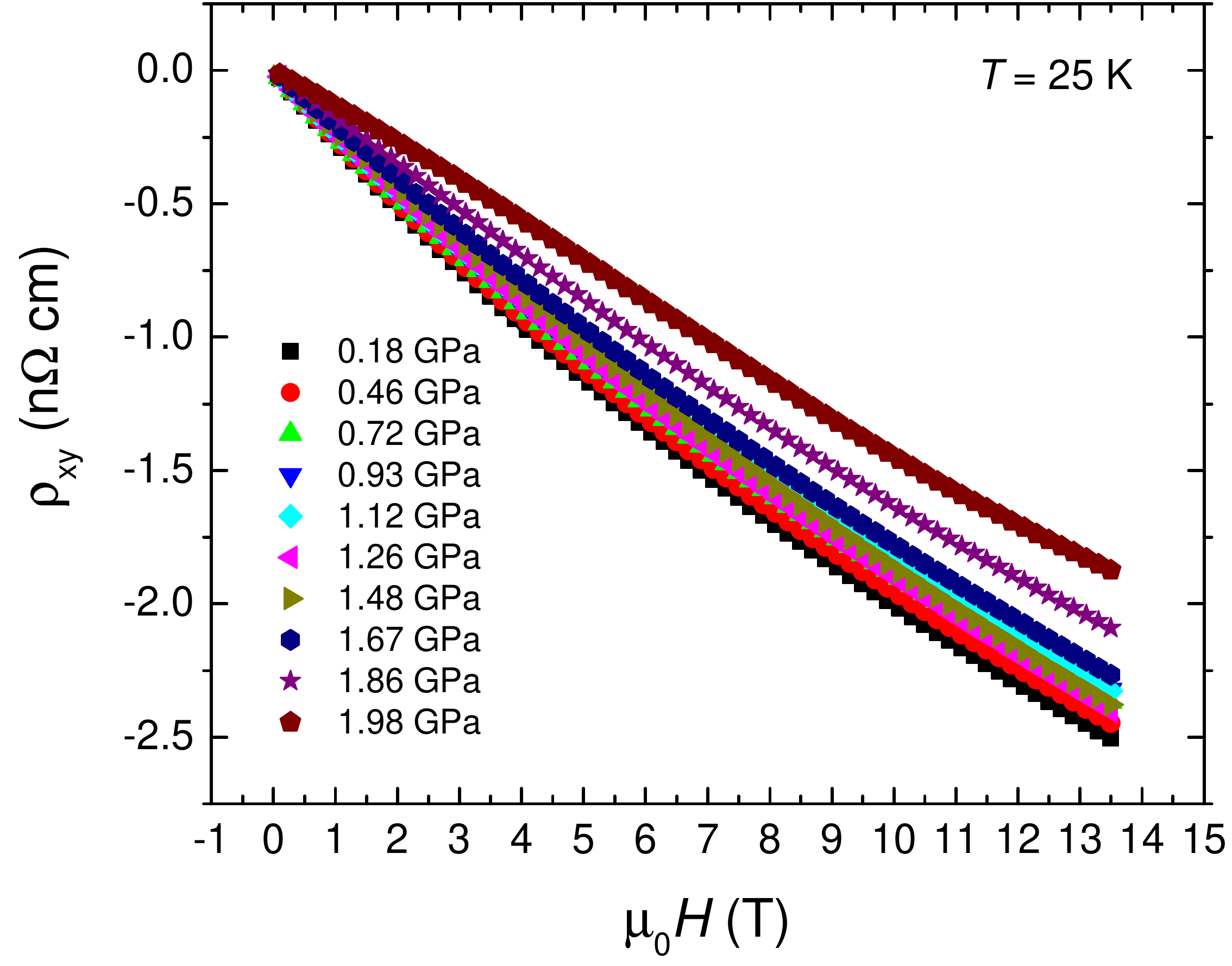} 
		\caption{Hall resistivity, $\rho_{xy}$, of Ba(Fe$_{1-x}$Co$_x$)$_2$As$_2$ ($x\,=\,0.02$) as a function of external magnetic field, $\mu_0 H$, at $T\,=\,25\,$K (a), 50\,K (b) and 125\,K (c) at different pressures. }
		\label{fig:Hall-raw}
		\end{center}
		\end{figure}	
	
	\subsection{Results}
		
	Figure \ref{fig:Hall-raw} shows Hall resistivity, $\rho_{xy}$, of Ba(Fe$_{1-x}$Co$_x$)$_2$As$_2$ ($x\,=\,0.02$) as a function of external magnetic field, $\mu_0 H$, at different pressures 0.18\,GPa$\,\le\,p\,\le\,$1.98\,GPa. All data were collected at three different temperatures, $T\,=\,25\,$K, 50\,K and 125\,K. Whereas the data collected at 25\,K and 50\,K correspond to the Hall effect in the antiferromagnetic state at $T\,<\,T_N$ at all pressures, the data at 125\,K is taken in the paramagnetic, tetragonal state at all pressures. At low fields, $\rho_{xy}$ exhibits an almost $H$-linear behavior with deviations occurring at higher fields, likely due to the multi-band nature of the iron-pnictide materials and the impact of magnetic order on the Fermi surface.
	
		\begin{figure}[h!]
		\begin{center}
		\includegraphics[width=\columnwidth]{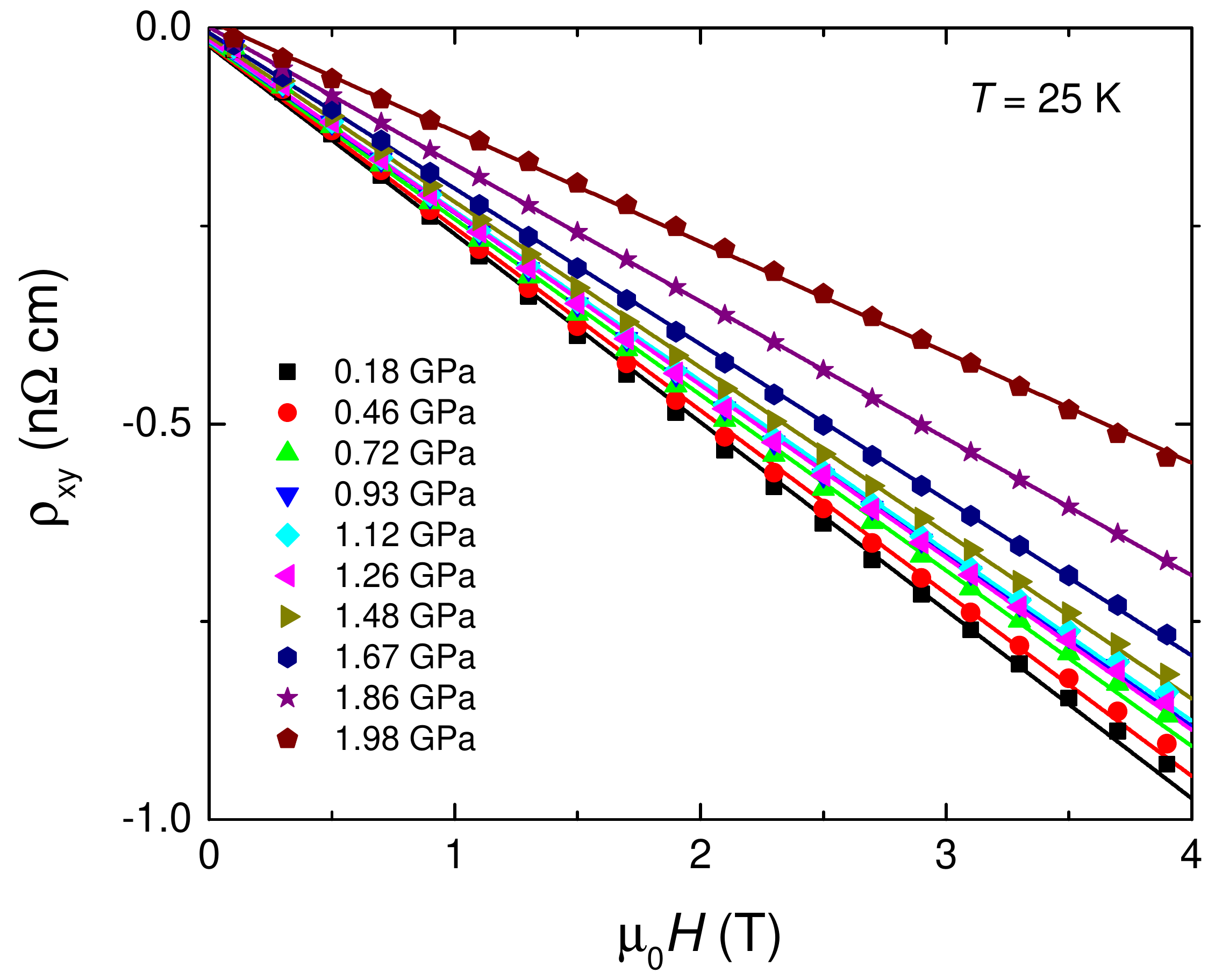} 
		\caption{Hall resistivity, $\rho_{xy}$, (symbols) of Ba(Fe$_{1-x}$Co$_x$)$_2$As$_2$ ($x\,=\,0.02$) as a function of external magnetic field, $\mu_0 H$, at $T\,=\,25\,$K in the low-field region up to 4\,T. Straight lines correspond to linear fits to the $\rho(H)$ data sets.}
		\label{fig:Hall-raw-blowup}
		\end{center}
		\end{figure}
		
\begin{figure}[h!]
		\begin{center}
		\includegraphics[width=\columnwidth]{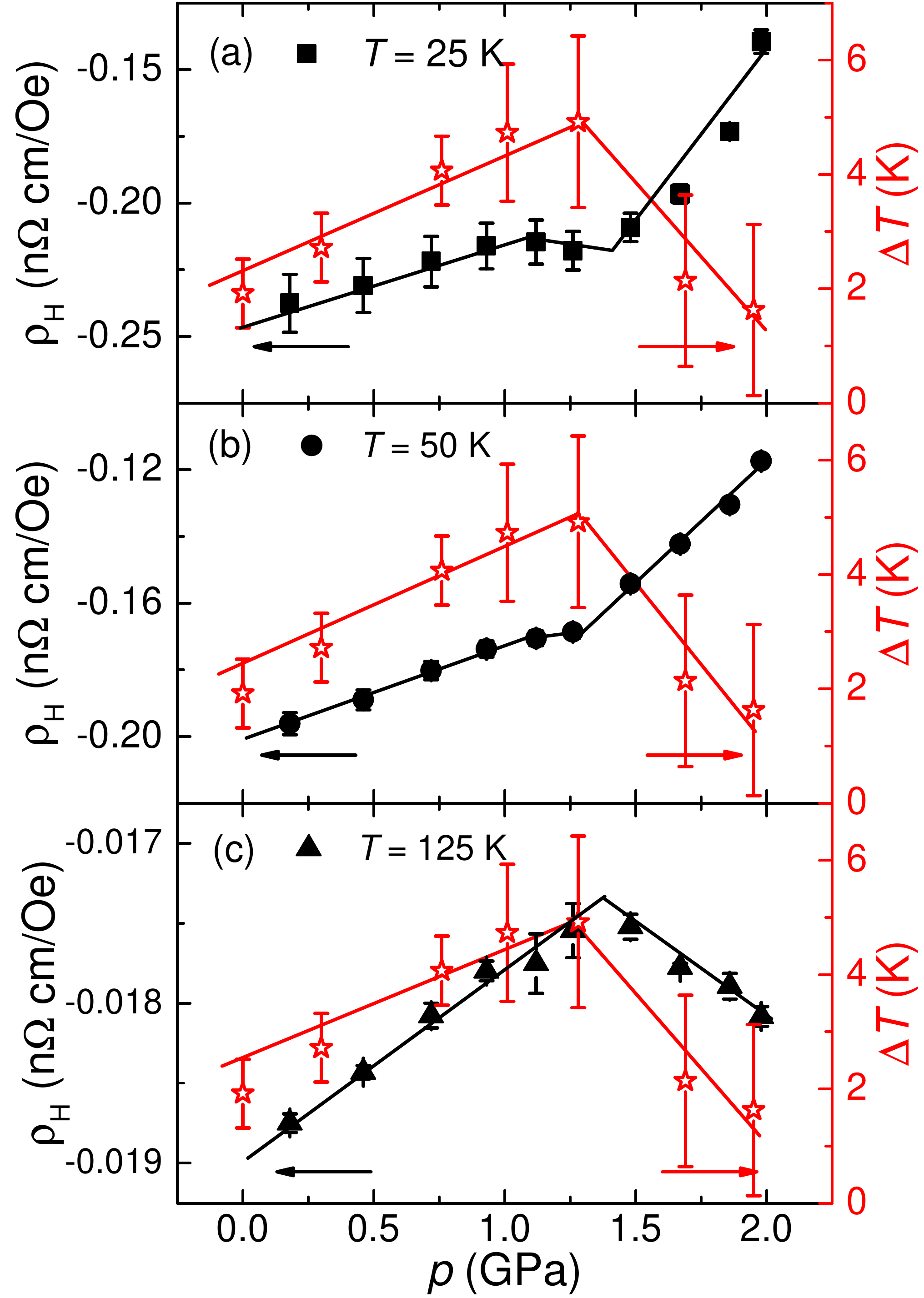} 
		\caption{Pressure-dependent Hall coefficient, $\rho_{H}$, (left axis, filled squares) taken at $T\,=\,$25\,K (top), 50\,K (middle) and 125 K (bottom) as well as $\Delta T\,=\,T_s-T_N$ (open stars) vs. $p$ for Ba(Fe$_{1-x}$Co$_x$)$_2$As$_2$ ($x\,=\,0.02$). Solid lines are guide to the eyes.}
		\label{fig:Hall-all}
		\end{center}
		\end{figure}

		The data presented in Fig.\,\ref{fig:Hall-raw} was used to extract the evolution of the Hall coefficient $\rho_{H}\,=\,\rho_{xy}/H$ as a function of $p$. To this end, we evaluated the slope of the $\rho$ data at low fields up to 4\,T. In this field range, $\rho(H)$ can be described to a good approximation by a linear behavior (see Fig.\,\ref{fig:Hall-raw-blowup} for a blow-up of the low-field region at $T\,=\,25\,$K). To estimate the error, which results from choosing this particular procedure, we fitted various low-field ranges of the $\rho_{H}$ vs. $H$ data (0\,T to 2\,T, 0\,T to 3\,T, 0\,T to 4\,T and 0\,T to 5\,T). The error bar for each $\rho_{H}$ data point in Fig.\,3 of the main manuscript and in Fig.\,\ref{fig:Hall-all} corresponds to the standard deviation of the extracted slopes of these various fits. The so-calculated errors are representing an upper boundary of the error bar, resulting from the analysis of our data.
		
		The evolution of this slope with pressure is compiled for all three temperatures in Fig.\,\ref{fig:Hall-all}. As clearly seen in this plot, a break of slope in $\rho_{H}$ vs. $p$ can  be observed at $p\,\,\approx\,1.3\,$GPa at all three temperatures investigated. This indicates that changes of the Fermi surface do not only occur in the antiferromagnetic state, but also in the paramagnetic, tetragonal state.
	
	\section{Resistance under pressure}	
	
	In the following, we will compare our result from specific heat measurements (i.e., a thermodynamic quantity) with measurements of resistance (i.e., a transport quantity) under pressure, as, in general, both quantities should display signatures of the magnetic and structural phase transition. As this family of compounds is known to be sensitive to non-hydrostatic pressure components (see e.g. Refs.\,\citen{Yamazaki10,Canfield09}), we omit a discussion of the existing literature data taken at higher pressures and therefore in inevitably less hydrostatic conditions\cite{Torikachvili15}. Instead, we present here a new data set up to 2\,GPa, which was collected in the same pressure environment as the specific heat data. This comparison supports our conclusions of a progressive splitting of $T_S$ and $T_N$, drawn in the main text.
		
			\subsection{Experimental Details}
			
			Resistance under pressure was measured in a four-point configuration with current directed along the $ab$ plane. Contacts were made using Epo-tek H20E silver epoxy. AC resistance was measured by a LakeShore 370 Resistance Bridge. Measurements of the resistance were performed in the same pressure cell (similar to the one described in Ref.\,\citen{Budko84}) as the specific heat measurements discussed in the main text.

			\subsection{Results on BaFe$_2$As$_2$}

		\begin{figure}[h!]
		\begin{center}
		\includegraphics[width=\columnwidth]{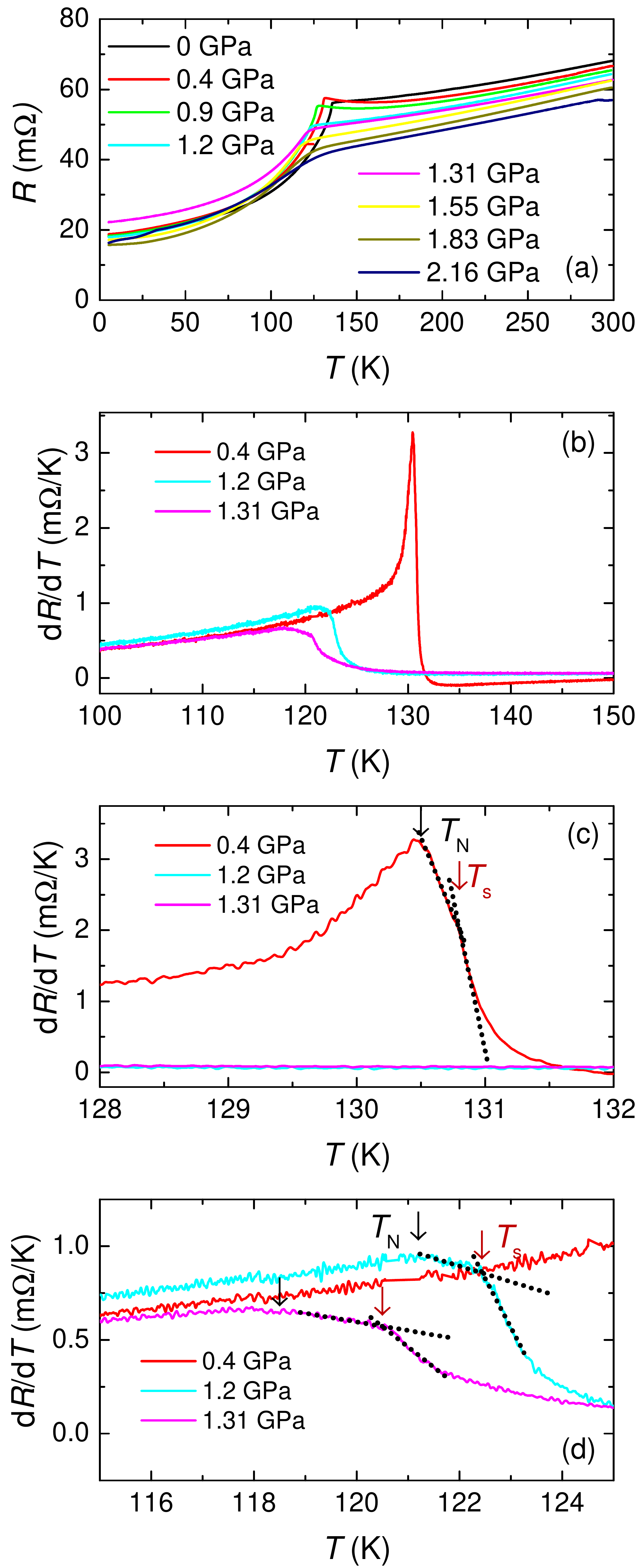} 
		\caption{(a) Resistance, $R$, of BaFe$_2$As$_2$ as a function of temperature, $T$, for different pressures (0 GPa\,$\le\,p\,\le\,$2.16\,GPa). Data are slightly offset with respect to each other for clarity; (b) First derivative of resistance data, d$R$/d$T$, for a few representative pressure values; (c) and (d) Blow-ups of the data shown in (b). Black (red) arrows indicate the position of the antiferromagnetic (structural) transition temperature at $T_N$ ($T_s$).}
		\label{fig:RTP}
		\end{center}
		\end{figure}
		
\begin{figure}[h!]
		\begin{center}
		\includegraphics[width=\columnwidth]{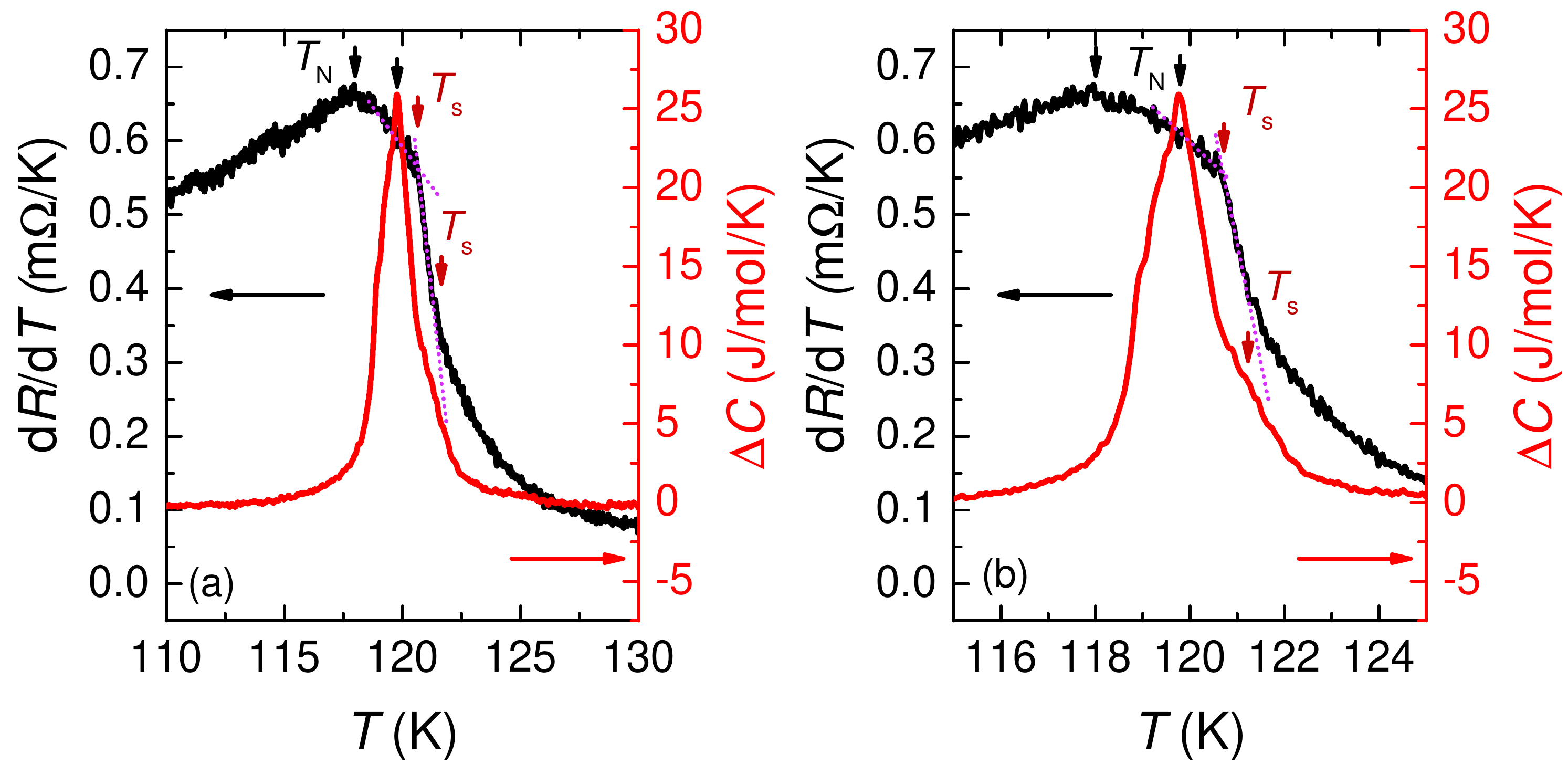} 
		\caption{(a) Comparison of derivative of resistance data, d$R$/d$T$, (left axis) and specific heat data, $\Delta C$, on BaFe$_2$As$_2$ at pressures of 1.31\,GPa; (b) Blow-up of the data shown in (a).}
		\label{fig:RvsC}
		\end{center}
		\end{figure}

		Figure \ref{fig:RTP} shows the collected data of the resistance, $R$, as a function of $T$ at different pressures up to 2.16\,GPa. All curves resemble a resistance behavior that is consistent with previous pressure studies\cite{Colombier09,Fukazawa08,Yamazaki10,Matsubayashi09,Ishikawa09}. The resistance shows metallic behavior in the entire temperature range and a pronounced downturn at a characteristic temperture, which is usually associated with the antiferromagnetic and structural transition temperatures $T_N$ and $T_S$. To identify the individual transition temperatures and define criteria, we show in Fig.\,\ref{fig:RTP}\,(b-d) the derivative of the resistance data, d$R$/d$T$, for a few selected pressure points which represent the characteristic evolution of features upon increasing pressure. At all pressures, d$R$/d$T$ displays a pronounced maximum (see black arrow in Fig.\,\ref{fig:RTP}), which we assign in the following to the antiferromagnetic transition temperature $T_N$, as done in previous works on BaFe$_2$As$_2$ as well as recent works on the Co-substituted BaFe$_2$As$_2$ (see e.g. Ref.\,\citen{Hristov18}). In the latter work\cite{Hristov18}, it was argued that the Fisher-Langer relation is applicable in the case of Ba(Fe$_{0.975}$Co$_{0.025}$)$_2$As$_2$ for the resistive feature at the antiferromagnetic and structural phase transitions. The validity of the Fisher-Langer relation implies $C(T)\,\propto\,$d$R(T)$/d$T$. When analyzing transport data and following the Fisher-Langer relation, a pronounced maximum at $T_N$ and a kink/shoulder at $T_S$ in d$R$/d$T$ can be expected. Indeed, we find a kink in d$R$/d$T$ at $T\,>\,T_N$ at all pressures (see red arrow) and assign this to $T_s$ by using the intersection point of two straight lines. In particular, the kink becomes clearly visible at higher pressures (1.2\,GPa and higher). In Fig.\,\ref{fig:RvsC}, we compare explicitly $C$ and d$R$/d$T$ data at the same pressure value. Even though this comparison shows that features in $C(T)$ and d$R(T)$/d$T$ are indeed similar, this figure also discloses slight differences in the absolute transition temperature values, which we assign to crystal-to-crystal variations in different batches.
		
\begin{figure}[h!]
		\begin{center}
		\includegraphics[width=\columnwidth]{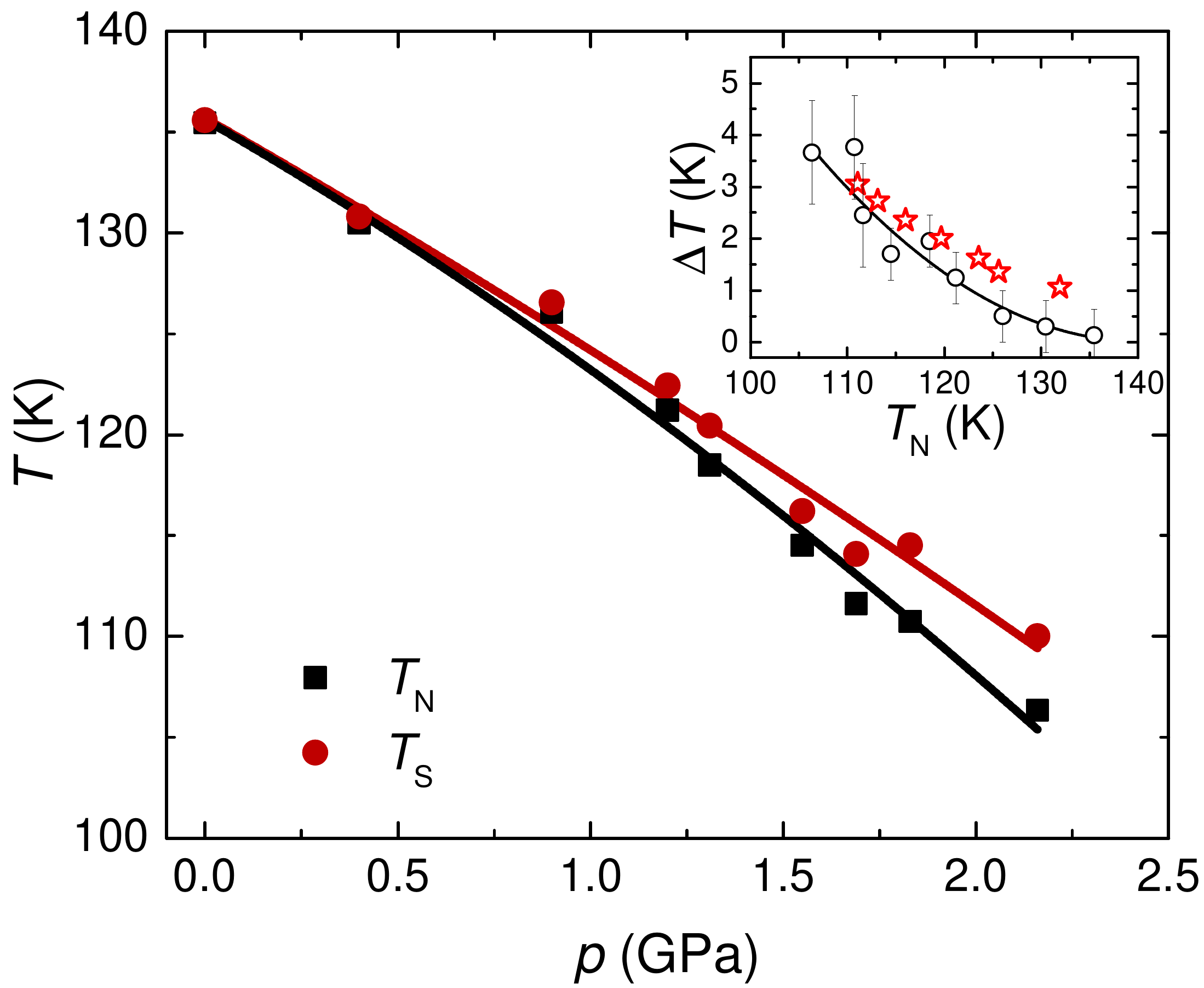} 
		\caption{Constructed temperature $T$-pressure $p$ phase diagram of BaFe$_2$As$_2$ showing the positions of antiferromagnetic and structural transition temperatures at $T_s$ and $T_N$. Inset: $\Delta T\,=\,T_s-T_N$ as a function of $T_N$, inferred from the resistance data (open black circles) and specific heat data (red stars).}
		\label{fig:RTP-PD}
		\end{center}
		\end{figure}

		The respective phase transition temperatures are compiled in a $T$-$p$ phase diagram, shown in Fig.\,\ref{fig:RTP-PD}. Both transitions ($T_S$ and $T_N$) are suppressed with $p$, however $T_S$ at a slower rate than $T_N$. Consequently, the splitting $\Delta T\,=\,T_S-T_N$ becomes larger upon increasing pressure. The splitting, inferred from the transport data, amounts to $\approx\,4\,$K at highest pressure of 2.16\,GPa. Compared to the evolution of $\Delta T$, inferred from specific heat measurements (see Inset), we find a very similar evolution of $\Delta T$ with $T_N(p)$. Therefore a careful analysis of transport data under pressure, taken in the same environment, confirm our conclusions drawn from specific heat measurements in the main text.

\bibliographystyle{apsrev}

\begin{thebibliography}{69}
\bibitem[{com()}]{comment1}
\bibinfo{note}{So far, the only study \cite{Wu13} which reports a study of
  magnetic and structural properties under pressure on the parent compound
  BaFe$_2$As$_2$ combined x-ray diffraction measurements with synchrotron
  M\"ossbauer spectroscopy (SMS). The SMS technique, as outlined in their
  manuscript, is sensitive to mesoscopic spin alignments, rather than the
  thermodynamic phase transition of the macroscopic spin system. The latter is
  the focus of the present work.}
\end{thebibliography}

\end{document}